\RequirePackage{etoolbox}
\csdef{input@path}{{style/}{graphics/}}
\documentclass[MSNbibl,nameyear,dvips]{arxstspdf}
\usepackage{graphicx}
\usepackage{url,breakurl}
\usepackage{flushend}
\usepackage{stfloats}

\volume{29}
\issue{4}
\pubyear{2014}
\firstpage{529}
\lastpage{558}
\doi{10.1214/13-STS453} 

\makeatletter
\renewcommand{\citep}[1]{\citeauthor{#1}, \citeyear{#1}}
\newcommand{\eqref}[1]{(\ref{#1})}
\makeatother

\begin{document}
\begin{frontmatter}

\title{Standardization and Control for Confounding in Observational
Studies: A~Historical Perspective}
\runtitle{Standardization: A Historical Perspective}

\begin{aug}
\author[A]{\fnms{Niels} \snm{Keiding}\corref{}\ead[label=e1]{nike@sund.ku.dk}}
\and
\author[B]{\fnms{David} \snm{Clayton}\ead[label=e2]{dc208@cam.ac.uk}}
\address[A]{Niels Keiding is  Professor, Department of Biostatistics,
University of Copenhagen, P.O. Box 2099, Copenhagen 1014, Denmark \printead{e1}.}
\address[B]{David Clayton is Emeritus Honorary Professor of Biostatistics,
University of Cambridge, Cambridge CB2 0XY, United Kingdom \printead{e2}.}
\affiliation{University of Copenhagen and University of Cambridge}
\runauthor{N. Keiding and D. Clayton}

\end{aug}

%
\begin{abstract}
Control for confounders in observational studies was
generally handled through stratification and standardization until the
1960s. Standardization typically reweights the stratum-specific rates
so that exposure categories become comparable.
With the development first of loglinear models, soon also of nonlinear
regression techniques (logistic regression, failure time regression)
that the emerging computers could handle, regression modelling became
the preferred approach, just as was already the case with multiple
regression analysis for continuous outcomes.
Since the mid 1990s it has become increasingly obvious that weighting
methods are still often useful, sometimes even necessary.
On this background we aim at describing the emergence of the modelling
approach and the refinement of the weighting approach for confounder
control.
\end{abstract}

%

\end{frontmatter}

\section{Introduction: Confounding and Standardization}
\label{sec:intro}
In this paper we survey the development of modern methods for
controlling for confounding in observational studies, with a primary
focus on discrete responses in demography, epidemiology and social
science. The forerunners of these methods are the methods of \emph
{standardization of rates}, which go back at least to the
18{th}
century [see \citet{Keiding-1987} for a review].
These methods tackle the problem of comparing rates
between populations with different age structures by applying
age-specific rates to a single ``target'' age structure and,
thereafter, comparing predicted \textit{marginal} summaries in this target
population. However, over the 20{th} century, the
methodological focus swung toward indices which summarize comparisons of
\textit{conditional} (covariate-specific) rates. This
difference of approach has, at its heart, the distinction between, for
example, a ratio of averages and an average of ratios---a
distinction discussed at some length in the important papers by
\citet{Yule-1934} and \citet{Kitagawa-1964}, which we shall discuss in
Section~\ref{sec:stand20}. The change of emphasis from a marginal to
conditional focus led eventually
to the modern dominance of the regression modelling approach
in these fields. \citeauthor{clayton:hills:93} [(\citeyear{clayton:hills:93}),
page 135] likened the two
approaches to the two
paradigms for dealing with extraneous variables in experimental
science, namely, (a) to make a marginal
comparison after ensuring, by \textit{randomization}, that the
distributions of such variables are
equal, and (b) to fix, or \textit{control}, such
influences and make comparisons conditional upon these fixed values.
In sections following, we shall chart how, in observational studies,
statistical approaches swung from the former to the latter. Finally, we
note that some recent methodological
developments have required a movement in the
reverse direction.

%
\begin{table}
\caption{Age standardization: some notation}
\label{tab:notation}
\tabcolsep=0pt
\begin{tabular*}{\columnwidth}{@{\extracolsep{\fill}}lcc@{}}
\hline
& {\textbf{Study population}} & {\textbf{Standard population}}\\
\hline
No. of individuals & $A_1\cdots A_k$ & $S_1 \cdots S_k$\\
Age distribution & $a_1\cdots a_k, \sum a_i=1$ &
$s_1\cdots s_k, \sum s_i=1$\\
Death rates & $\alpha_1 \cdots\alpha_k$ & $\lambda_1 \cdots\lambda
_k$\\
Actual no. of deaths & $\sum A_i\alpha_i$ & $\sum S_i \lambda_i$\\
Crude death rate & $\sum A_i\alpha_i / \sum A_i$ &
$\sum S_i\lambda_i / \sum S_i$ \\
\hline
\end{tabular*}
\end{table}

We shall start by recalling the basic
concepts of direct and indirect standardization in the simplest case
where a \textit{study population} is to be compared to a
\textit{standard population}.
Table~\ref{tab:notation} introduces some notation, where there are $k$
age groups.
In indirect
standardization, we apply the age-specific death rates
for the standard population to the age distribution
of the study, yielding the counterfactual number of deaths
in the study population if the rates had been the same as the standard
rates. The Standardized Mortality Ratio (SMR)
is the ratio between the observed number of deaths in
the study population to this ``expected'' number:
\[
\operatorname{SMR} = \sum A_i\alpha_i \big/\sum
A_i\lambda_i.
\]
Note that the numerator does not require knowledge of the age
distribution of deaths in the study group. This property has often
been useful.

In \textit{direct standardization} one calculates what
the marginal death rate would have been in
the study population if its age distribution had been the same as in the
standard population:
\begin{eqnarray*}
\mbox{(Direct) standardized rate} &= &\sum s_i
\alpha_i\\
& =& \sum S_i\alpha_i \big/
\sum S_i.
\end{eqnarray*}
This is sometimes expressed relative to the marginal rate in
the standard population---the \textit{Comparative Mortality
Figure} (CMF):
\[
\operatorname{CMF} = \sum S_i\alpha_i \big/ \sum
S_i\lambda_i.
\]

\citet{Sato-Matsuyama-2003} and \citet{Hernan-Robins-2006}
gave concise and readable accounts of the connection of
standardization to modern causal analysis. Assume that, as in the
above simple situation, outcome is binary (death)
and exposure is binary---individuals are either
exposed (study population) or unexposed (standard population).
Each individual may be
thought of as having a different risk for each exposure state, even
though only one state can be observed in practice. In addition to
depending on exposure, risks depend
on a discrete confounder (age group).
The causal effect of the exposure can be defined as the ratio of
the marginal risk in a population of individuals had they been exposed
to the risk for the same individuals had they not been exposed.
Conditional
exchangeability is assumed; for a given value of the confounder
(in the present case, within each age group), the counterfactual risks
for each individual do not depend on the actual exposure status.
Then the marginal death rate in the unexposed (standard) population
of individuals had they been exposed
is estimated by the directly standardized rate, so that
the causal risk
ratio for the unexposed population is estimated by the CMF. Similarly, the
causal risk ratio for the exposed population
is estimated by the SMR. We may
estimate the death rate of the exposed population had they not been
exposed by the \textit{indirectly standardized death rate},
obtained by multiplying the crude rate in the standard
population by the SMR:
\[
\mbox{Indirect standardized rate} =
\frac{\sum S_i\lambda_i}{\sum S_i} \times\frac{\sum A_i\alpha_i}{\sum
A_i\lambda_i}.
\]
Both direct and indirect approaches are based on comparison of marginal
risks, although, as pointed out by \citet{Miettinen-1972}, they
focus on different ``target'' populations;
indirect standardization may be said to have the
study population as its target, while direct standardization
has the standard population as its target. Indeed,
the CMF is identical to the (reciprocal of)
the SMR if ``study'' and ``standard'' populations are interchanged.

In many epidemiological and biostatistical contexts it
is natural to use the total population (exposed${} + {}$unexposed) as basis
for statements about causal risk ratios. With $N_i = A_i + S_i$, the total
population size in age group $i$, the causal risk ratio in the total
population will be
\[
\frac{\sum N_i\alpha_i}{\sum N_i\lambda_i} =
\frac{\sum A_i (N_i/A_i) \alpha_i}{\sum S_i (N_i/S_i) \lambda_i}.
\]
This rearrangement of the formula shows that we may interpret
standardization with the total population as target as an inverse
probability weighting method in which the weighting compensates for
nonobservation of the counterfactual exposure state for each subject.
In the numerator, the contributions of the $A_i$ exposed subjects are
inversely weighted by $A_i/N_i$, which estimates the probability that
a subject in age group $i$ of the total study was observed in the
exposed state. Similarly, in the denominator, the $S_i$ unexposed
subjects are inversely weighted by the probability that a subject was
observed in the unexposed state. The method of inverse
probability weighting is an important tool in marginal structural
models and other methods in modern causal analysis.

Thus, while there are obvious similarities between direct and indirect
standardization, there are also important differences. In particular,
when the
aim is to compare rates in \textit{several} study populations, reversal
of the roles of study and standard population is no longer
possible and \citet{Yule-1934} pointed out important faults with the
indirect approach in this context. Such considerations will lead us,
eventually, to see indirect standardization as dependent on an
implicit model and, therefore, as a
forerunner of the modern conditional modelling approach.

The plan of this paper is to present selected highlights from the
historical development of confounder control with focus on the
interplay between marginal or conditional choice of target, on the one
hand, and the role of (parametric or nonparametric) statistical
models on the other. Section~\ref{sec:mortality}
recalls the development of standardization
techniques during the 19{th}
century. Section~\ref{sec:association} deals with early
20{th} century approaches to the problem of causal inference,
focusing particularly on the contributions of Yule and Pearson.
Section~\ref{sec:stand20} records highlights from the parallel
development in the social sciences, focusing on the further
development of standardization methods in the
20{th} century---largely in the social sciences.
Section~\ref{sec:cc} deals with the important developments in the 1950s
and early 1960s surrounding the analysis of the $2\times2\times K$
contingency table, and Section~\ref{sec:models} briefly summarizes
the subsequent rise and dominance of regression models.
Section~\ref{sec:transport} points out that the values of
parameters in (conditional) probability models are not always the only
focus of
analysis, that marginal predictions in different target populations
are often important, and that such predictions require careful
examination of our assumptions. Finally, Section~\ref{sec:conclusion}
contains a brief concluding summary.

Here we have used the word ``rate'' as a synonym for
``proportion'', reflecting usage at the time. It was later recognized that
a distinction should properly be made (\citep{elandt-johnson:75},
\citep{miettinen:76})
and modern usage reflects this. However, for this historical review it has
been more convenient to follow the older terminology.

\section{Standardization of Mortality Rates in the
19{th} century}
\label{sec:mortality}
\subsection*{Neison's Sanatory Comparison of Districts}
It is fair to start the description of direct and indirect
standardization with
the paper by \citet{Neison-1844}, read to the Statistical Society of
London on 15 January
1844, responding to claims made at the previous meeting (18 December
1843) of
the Society by \citet{Chadwick-1844} about ``representing the duration
of life''.

Chadwick was concerned with comparing mortality ``amongst different
classes of
the community, and amongst the populations of different districts and
countries''.
He began his article by quoting the 18{th} century
practice of using ``proportions of death''
(what we would now call the crude death rate): the simple ratio of
number of deaths in a year to the size of the
population that year. Under the Enlightenment age assumption of
stationary population,
it is an elementary demographic fact that the crude death rate is the
inverse of
the average life time in the population, but as Chadwick pointed out,
the stationarity
assumption was not valid in England at the time. Instead, Chadwick
proposed the
average age of death (i.e., among those dying in the year
studied). Neison responded:

\begin{quote}
That the average age of those who die in one community cannot be taken
as a
test of the value of life when compared with that in another district
is evident
from the fact that no two districts or places are under the same
distribution of
population as to ages.
\end{quote}

To remedy this, Neison proposed to not only calculate the average age
at death in
each district, but

\begin{quote}
also what would have been the average age at death if placed under the same
population as the metropolis.
\end{quote}

This is what we now call \textit{direct standardization}, referring the
age-specific
mortality rates in the various districts to the same age distribution.
A little later Neison remarked that

\begin{quote}
Another method of viewing this question would be to apply the same rate
of mortality
to different populations,
\end{quote}

\noindent what we today call \textit{indirect standardization}.

\citet{Keiding-1987} described the prehistory of indirect
standardization in 18{th}
century actuarial contexts; although Neison was himself an actuary, we
have found
no evidence that this literature was known to Neison, who apparently developed
direct as well as indirect standardization over Christmas
1843. \citeauthor{Schweber-2001} (\citeyear{Schweber-2001}, \citeyear{Schweber-2006}) [cf. \citet{Bellhouse-2008}]
attempted a historical--sociological discussion of the debate between
Chadwick and Neison.

%
\begin{table*}[t]
\caption{Distribution of deaths of Danish medical doctors 1815--1870,
as well as
the expected number of deaths if the doctors had been subjected to the
mortality
of the general (male) population, based on age-specific mortality
rates for Denmark as a whole as
well as on age-specific mortality rates separately for each of the
three districts Copenhagen, Provincial Towns, Rural Districts
[Westergaard (\citeyear{Westergaard-1882}), page 40]}\label{tab:tab1}
\begin{tabular*}{\textwidth}{@{\extracolsep{4in minus 4in}}lcccc@{}}
\hline
&&&\multicolumn{2}{c@{}}{\textbf{Expected number of deaths according
to}}\\
\cline{4-5}
&\textbf{Years at risk}&\textbf{Dead}& \textbf{three special districts} &
\multicolumn{1}{c@{}}{\textbf{whole country}}\\
\hline
Copenhagen & \phantom{\mbox{,}0}7127\phantom{0.} & 108 & 156 & \phantom{0}98\\
Provincial towns & \phantom{\mbox{,}0}9556.5 & 159 & 183 & 143\\
Rural districts & \phantom{\mbox{,}0}4213.5 & \phantom{0}74 & \phantom{0}53 &
\phantom{0}60\\[3pt]
Whole country & 20\mbox{,}897.0 & 341 & 392 &301\\
\hline
\end{tabular*}
\end{table*}

A few years later \citet{Neison-1851} published an elaborate survey
``On the rate of
mortality among persons of intemperate habits'' in which he wrote in
the typical
style of the time:

\begin{quote}
From the rate of sixteen upwards, it will be seen that the rate of mortality
exceeds that of the general population of England and Wales. In the
6111.5 years
of life to which the observations extend, 357 deaths have taken place;
but if
these lives had been subject to the same rate of mortality as the
population generally,
the number of deaths would only have been 110, showing a difference of
3.25 times.
\ldots If there be anything, therefore, in the usages of society
calculated to destroy
life, the most powerful is certainly the use of strong drink.
\end{quote}

In other words, an SMR of 3.25.

Expected numbers of deaths (indirect standardization) were calculated
in the English
official statistical literature, particularly by W. Farr,
for example, \citet{Farr-1859}, who
chose the standard mortality rates as the annual age-specific death
rates for 1849--1853
in the ``healthy districts'', defined as those with average crude
mortality rates
of at most 17$/$1000 [see \citet{Keiding-1987} for an example].
W. Ogle initiated routine
use of (direct)
standardization in the Registrar-General's
report of 1883,
using the 1881 population census of England and Wales as the
standard. In 1883,
direct standardization of official mortality statistics was also
started in Hamburg by G. Koch. Elaborate discussions on
the best choice of an international standard age distribution took
place over several biennial sessions of the International Statistical
Institute; cf. \citet{Korosi-1892-1893}, \citet{Ogle-1892} and
\citet{von-Bortkiewicz-1904}.

\subsection*{Westergaard and Indirect Standardization}

Little methodological refinement of the standardization methods seems
to have taken
place in the 19{th} century. One exception is the work
by the Danish
economist and statistician H. Westergaard, who already in his first
major publication,
\citet{Westergaard-1882} (an extension, in German, of a prize paper that
he had submitted
to the University of Copenhagen the year before), carefully described
what he called
\textit{die Methode der erwartungsm\"{a}ssig Gestorbenen} (the method
of expected
deaths), that is, indirect standardization. He was well aware of the
danger that other
factors could distort the result from a standardization by age alone
and illustrated
in a small introductory example the importance of what we would
nowadays call confounder
control, and how the method of expected number of deaths could be used
in this
connection.

Table~\ref{tab:tab1} shows that when comparing the mortality of medical
doctors with that of
the general population, it makes a big difference whether the
calculation of expected
number of deaths is performed for the country as a whole or
specifically (we would
say ``conditionally'') for each urbanization stratum. In Westergaard's words,
our English translation:

\begin{quote}
It is seen from this how difficult it is to conduct a scientific statistical
calculation. The two methods both look correct, and still yield very different
results. According to one method one would conclude that the medical
professionals
live under very unhealthy conditions, according to the other, that
their health
is relatively good.

The difficulty derives from the fact that there \textit{exist two
causes}: the
medical profession and the place of residence; both causes have to be
taken into
account, and if one neglects one of them, the place of residence, and
only with
the help of the general life table considers the influence of the
other, one will
make an erroneous conclusion.

The safest is to continue the stratification of the material until no
further disruptive
causes exist; if one has no other proof, then a safe sign that this
has been achieved,
is that further stratification of the material does not change the results.
\end{quote}

This general strategy of stratifying until the theoretical variance
had been achieved, eliminating any residual heterogeneity beyond the
basic binomial variation, was heavily influenced by the then current
attempts by Quetelet and Lexis in identifying homogeneous subgroups in
data from social statistics, for which the normal distribution could
be used, preferably with the interpretation of an approximation to the
binomial [see \citet{Stigler-1986} for an exposition on Quetelet and
Lexis]. In his review of the book, \citet{Thiele-1881} criticized
Westergaard's account for overinterpreting the role of mathematical
results such as the law of large numbers (as the central limit theorem
was then termed) in empirical sciences. As we shall see, however,
Westergaard remained fascinated by the occurrence of binomially
distributed data in social statistics.

Westergaard also outlined a derivation of the standard error of the
expected number of deaths, using what we would call a Poisson
approximation argument similar to the famous approximation by
\citet{Yule-1934} fifty years later for the standard error of the SMR.
We shall
see later that \citet{Kilpatrick-1962} had the last word on this matter by
justifying Yule's approximation in the framework of maximum likelihood
estimation in a proportional hazards model.

Standard error considerations accompany the many concrete calculations
on human mortality throughout Westergaard's book from 1882, which in
our view is original in its efforts to integrate statistical
considerations of uncertainty into mortality analysis, with indirect
standardization as the central tool. In the second edition of the book,
\citeauthor{Westergaard-1901} [(\citeyear{Westergaard-1901}), page 25]
explained that the method of expected number
of deaths has (our translation)

\begin{quote}
the advantage of summarizing many small series of observations with
all their random differences without having to abandon the
classification according to age or other groupings (e.g., occupation,
residence etc.), in other words obtaining the advantage of an
extensive material, without having to fear its disadvantages.
\end{quote}

When \citet{Westergaard-1916} finally presented his  views on
statistics in English, the printed comments in what we now call JASA
were supplemented by a detailed review by \citet{Edgeworth-1917} for the
Royal Statistical Society. \citeauthor{Westergaard-1916}
[(\citeyear{Westergaard-1916}), page 246] had gone as far
as to write:

\begin{quote}
In vital or economic statistics most numbers have a much wider
margin of deviation
than is experienced in games. Thus the death rate, the birth rate, the marriage
rate, or the relative frequency of suicide fluctuates within wide
limits. But it
can be proved that, by dividing the observations, sooner or later a
marked tendency
to the binomial law is revealed in some parts of the
observations. Thus, the birth
rate varies greatly from year to year; but every year nearly the same
ratio between
boys and girls, and the same proportions of stillbirths, and of twins
are observed \ldots
\end{quote}

\noindent and (page 248)

\begin{quote}
\ldots there is no difficulty in getting several important results concerning
relative numbers. The level of mortality may be very different from
year to year,
but we can perceive a tendency to the binomial law in the relative
numbers, the
death rates by age, sex, occupation etc.
\end{quote}

Edgeworth questioned that ``Westergaard's pa\-nacea'' would work as a general
remedy in all situations, and continued:

\begin{quote}
It never seems to have occurred to him that the ``physical'' as distinguished
from the ``combinatorial'' distribution, to use Lexis' distinction, may be
treated by the law of error [the normal distribution].
\end{quote}

Edgeworth here referred to the empirical (physical) variance as
opposed to the binomial (combinatorial). \citet{Lexis-1876}, in the
context of time series of rates, had defined what
we now call the overdispersion ratio between these two.

\subsubsection*{Indirect standardization does not require the age
distribution of the cases}
Regarding standardization, \citeauthor{Westergaard-1916}
[(\citeyear{Westergaard-1916}), page 261 ff.]
explained and exemplified
the method of expected number of deaths, as usual without quoting
Neison or other
earlier users of that method, such as Farr, and went on:

\begin{quote}
English statisticians often use a modification of the method just
described of
calculating expected deaths; viz., the method of ``standards'' (in fact
the method
of expected deaths can quite as well claim the name of a ``standard'' method),
\end{quote}

\noindent and after having outlined direct standardization concluded,

\begin{quote}
In the present case the two forms of comparison lead to nearly the
same result,
and this will generally be the case, if the age distribution in the
special group
is not much different from that of the general population. But on the
whole the
method described last is a little more complicated than the
calculation of expected
deaths, and in particular not applicable, if the age distribution of
the deaths
of the barristers and solicitors is unknown.
\end{quote}

This last point (that indirect standardization does not require the
breakdown of
cases in the study population by age) has often been emphasized as an
important advantage of indirect standardization.
An interesting application was the study of the emerging \textit{fall
of the birth
rate} read to the Royal Statistical Society in December 1905 by
\citet{Newsholme-Stevenson-1906} and \citet{Yule-1906}.
[\citet{Yule-1920} later presented a concise popular version
of the main findings to the Cambridge Eugenics Society, still
interesting reading.]
The problem was that English birth statistics did not include the age
distribution
of the mother, and it was therefore recommended to use some standard
age-specific
birth rates (here: those of Sweden for 1891) and then indirect standardization.

\subsection*{Westergaard and an Early Randomised Clinical Trial}

Westergaard (\citeyear{Westergaard-1918}) published a lengthy rebuttal
(``On the future of statistics'') to Edgeworth's critique.
Westergaard was here mainly concerned with the statistician's overall ambition
of contributing
to ``find the causality'', and with a main point being his criticism
of ``correlation based on Bravais's formula'' as not indicating causality.
However, he also had an interesting, albeit somewhat cryptic,
reference to a topic
that was to become absolutely central in the coming years: that simple binomial
variation is justified under random sampling. In his 1916 paper, he
had advocated
(page 238) that

\begin{quote}
in many cases it will be practically impossible to do without representative
statistics.
\end{quote}

%
\begin{table*}[b]
\tabcolsep=0pt
\caption{Yule's analysis of the association between smallpox
vaccination and attack rates (defined as percentage contracting
the disease in ``invaded~household'')}\label{tab:yule}
\begin{tabular*}{\textwidth}{@{\extracolsep{4in minus 4in}}lccccccc@{}}
\hline
&& \multicolumn{2}{c}{\textbf{Attack rate under 10}} &
\multicolumn{2}{c}{\textbf{Attack rate over 10}} &
\multicolumn{2}{c}{\textbf{Yule's} $\bolds{Q}$}\\
\ccline{3-4,5-6,7-8}
\textbf{Town} & \textbf{Date} & \textbf{Vaccinated} & \textbf
{Unvaccinated} & \textbf{Vaccinated} &
\textbf{Unvaccinated} & $\bolds{<\!10}$ & \multicolumn{1}{c@{}}{$\bolds
{>\!10}$}\\
\hline
Sheffield & 1887--1888 & \phantom{0}7.9 & 67.6 & 28.3 & 53.6 & 0.92 &
0.49\\
Warrington & 1892--1893 & \phantom{0}4.4 & 54.5 & 29.9 & 57.6 & 0.93 &
0.52\\
Dewsbury & 1891--1892 & 10.2 & 50.8 & 27.7 & 53.4 & 0.80 & 0.50\\
Leicester & 1892--1893 & \phantom{0}2.5 & 35.3 & 22.2 & 47.0& 0.91 &
0.51\\
Gloucester & 1895--1896 & \phantom{0}8.8 & 46.3 & 32.2 & 50.0 & 0.80 &
0.36\\
\hline
\end{tabular*}
\end{table*}

[\citet{Edgeworth-1917} taught Westergaard that the correct phrase was
``sampling'',
and Westergaard replied that English was for him a foreign language.]
To illustrate
this, \citeauthor{Westergaard-1916} [(\citeyear{Westergaard-1916}),
page 245] wrote:

\begin{quote}
The same formula in a little more complicated form can be applied to
the chief
problem in medical statistics; viz., to find whether a particular
method of treatment
of disease is effective. Let the mortality of patients suffering from
the disease
be $p_2$, when treated with a serum, $p_1$, when treated without
it, and let the numbers in each case be $n_2$ and $n_1$. Then the
mean error of the difference between the frequencies of dying in the
two groups
will be $\sqrt{p_1q_1/n_1 + p_2q_2/n_2}$
and we can get an approximation by putting the observed relative values instead
of $p_1$ and $p_2$.
\end{quote}

In his rebuttal, \citeauthor{Westergaard-1918}
[(\citeyear{Westergaard-1918}), page 508] revealed that this
was not just a hypothetical example:

\begin{quote}
A very interesting method of sampling was tried several years ago
in a Danish hospital
for epidemic diseases in order to test the influence of serum on
patients suffering
from diphtheria. Patients brought into the hospital one day were
treated with serum,
the next day's patients got no injection, and so on alternately. Here
in all
probability the two series of observations were homogeneous.
\end{quote}

Westergaard here referred to the experiment by \citet{Fibiger-1898},
discussed by \citet{Hrobjartsson-etal.-1998}, as
``the first randomized clinical trial'' and further documented
in the James Lind Library:
\href{http://www.jameslindlibrary.org/illustrating/records/om-serumbehandling-af-difteri-on-treatment-of-diphtheria-with-s/key\_passages}{http://www.jameslindlibrary.org/illustrating/}\break
\href{http://www.jameslindlibrary.org/illustrating/records/om-serumbehandling-af-difteri-on-treatment-of-diphtheria-with-s/key\_passages}{records/om-serumbehandling-af-difteri-on-}\break
\href{http://www.jameslindlibrary.org/illustrating/records/om-serumbehandling-af-difteri-on-treatment-of-diphtheria-with-s/key\_passages}{treatment-of-diphtheria-with-s/key\_passages}.

\section{Association, and Causality: Yule, Pearson and Following}
\label{sec:association}

The topic of causality in the early statistical literature is
particularly associated with Yule and with Pearson, although they were
far from the first to grapple with the problem. Yule
considered the topic mainly in the context of discrete data, while
Pearson considered mainly continuous variables. It is perhaps this
which led to some dispute between them, particularly in regard to
measures of association. For a detailed review of their differences,
see \citet{Aldrich-1995}.

\subsection*{Yule's Measures of Association and Partial Association}

For a $2\times2$ table with entries $a, b, c, d$, \citet{Yule-1900} defined
the association measure $Q = (ad - bc)/(ad + bc)$, noting that it
equals 0 under independence and 1 or $-1$ under complete association.
There are of course many choices of association measure that fulfil
these conditions. \citeauthor{Pearson-1900} [(\citeyear{Pearson-1900}),
pages 14--18] immediately made strong
objections to Yule's choice; he wanted a parameter that agreed well
with the correlation if the $2\times2$ table was generated from an
underlying bivariate
normal distribution. The discussion between Yule and Pearson and
their camps went on for more than a decade. It was chronicled from a
historical--sociological viewpoint by MacKenzie
(\citeauthor{MacKenzie-1978}, \citeyear{MacKenzie-1978,MacKenzie-1981}).

That he regarded the concrete values of $Q$ meaningful outside of 0 or 1
is illustrated by his analysis of the association between smallpox
vaccination and attack, as measured by $Q$, in several
towns (Table~\ref{tab:yule}). The values of $Q$ were much higher for
young children than for
older people, but did not vary markedly between different towns,
despite considerable variation in attack rates.
This use of $Q$
is different from an immediately interpretable population summary
measure and it is closer to how we use models and parameters
today. Indeed, since $Q$ is a simple
transformation of the odds ratio, $(ad)/(bc)$, Yule's analyses
of association anticipate modern orthodoxy ($Q = 0.9$ corresponds to
an odds ratio
of 19, and $Q=0.5$ to an odds ratio of 3).

Yule's view on \textit{causal} association was largely expounded by
consideration of its antithesis, which he termed ``illusory'' or
``misleading'' association. Chief amongst the reasons for such
noncausal association he identified as that due to the direct effect of
a third variable on outcome. His discussion of this phenomenon in
\citet{Yule-1903} (under the heading ``On the fallacies that may be
caused by the mixing of distinct records'') and, later, in his 1911 book
(\citep{Yule-1911}) came to be termed ``Yule's paradox'',
describing the situation in
which two variables are \textit{marginally} associated but not associated
when examined in subgroups in which the third, causal, variable is
held constant.
The idea of measuring the strength of association
holding further variables constant, which Yule termed ``partial''
association, was thus identified as an important protection against
fallacious causal explanations. However, he did not formally consider
modelling these partial associations. Indeed, he commented
(\citep{Yule-1900}):

\begin{quote}
The number of possible partial coefficients becomes very high as soon
as we
go beyond four or five variables.
\end{quote}

Yule did not discuss more parsimonious definitions of
partial association, although clearly he regarded the empirical
stability of $Q$ over different subgroups of data as a strong
point in its favour. Commenting on some data on recovery from
smallpox, in \citet{Yule-1912}, he later wrote:

\begin{quote}
This, as it seems to me, is a most important property \ldots
If you told any man of ordinary intelligence that the association between
treatment and recovery was low at the beginning of the experiment,
reached a maximum when 50 per cent. of the cases were
treated and then fell off again as the proportion of cases treated
was further increased, he would, I think, be legitimately puzzled,
and would require a good deal of explanation as to what you meant
by association. \ldots The association
coefficient $Q$ keeps the same value throughout, quite unaffected
by the ratio of cases treated to cases untreated.
\end{quote}

\subsection*{Pearson and Tocher's Test for Identity of Two Mortality
Distributions}

Pearson regarded the theory of correlation as of fundamental
importance, even to the extent of replacing ``the old idea of
causality'' (\citep{Pearson-1910}). Nevertheless, he recognised the
existence of ``spurious'' correlations due to incorrect use of indices
or, later, due to a third variable such as race
(\citep{Pearson-Lee-Bramley-Moore-1899}).

Although most of Pearson's work concerned correlation between
continuous variables, perhaps the most relevant to our present
discussion is his work, with J.~F. Tocher, on comparing mortality
distributions. \citet{Pearson-Tocher-1916} posed the question of finding
a proper test for comparing two mortality distributions. Having
pointed out the problems of comparing crude mortality rates, they
considered comparison of standardized rates (or, rather, proportions).
In their notation, if we denote the number of deaths in age group $s$
($=1, \ldots, S$)
in the two samples to be compared by $d_s, d_s^\prime$ and the
corresponding numbers of persons at risk by $a_s, a_s^\prime$, then two
age-standardized rates can be calculated as
\[
M = \frac{1}{A} \sum A_s \frac{d_s}{a_s}
\quad\mbox{and} \quad M^\prime= \frac{1}{A} \sum
A_s \frac{d_s^\prime}{a_s^\prime},
\]
where $A_s$ represent the standard population in age group $s$ and
$A = \sum A_s$. Noting that the difference between standardized
rates can be expressed as a weighted mean of the differences between
age-specific rates,
\[
M^\prime- M = \sum\frac{A_s}{A} \biggl( \frac{d_s}{a_s}
- \frac{d_s^\prime}{a_s^\prime} \biggr),
\]
they showed that, under the null hypothesis that the true rates
are equal for the two groups to be compared,
\[
\operatorname{Var}\bigl(M^\prime- M\bigr) = \sum\biggl(
\frac{A_s}{A} \biggr)^2 p_s(1-p_s)
\biggl(\frac{1}{a_s} + \frac{1}{a_s^\prime} \biggr),
\]
where $p_s$ denote the (common) age-specific binomial
probabilities. Finally, for large studies, they advocated estimation
of $p_s$ by $(d_s+d_s^\prime)/(a_s+a_s^\prime)$ and treating
$(M^\prime-M)$ as approximately normally distributed or, equivalently,
\[
Q^2 = \frac{(M^\prime-M)^2}{\widehat{\operatorname{Var}}(M^\prime-M)}
\]
as a chi-squared variate on one degree of freedom (note that
their $Q^2$ is not
directly related to Yule's $Q$). However,
they pointed out a major problem with this approach; that
different choices of standard population lead to different answers, and
that there would usually be
objections to any one choice. In an attempt to resolve this difficulty,
they
proposed choosing the weights $A_s/A$ to maximise the test statistic
and showed that the resulting $Q^2$ is a $\chi^2$ test on $S$ degrees
of freedom. This is because, as \citet{Fisher-1922} remarked,
each age-specific $2\times2$-table of districts
vs. survival contributes an independent degree of freedom to the $\chi^2$
test.

Pearson and Tocher's derivation of this test anticipates the
much later, and more general, derivation of the score test as a
``Lagrange multiplier test''.
However, the maximized test statistic
could sometimes involve negative weights, $A_s$, which they described as
``irrational''. This feature of the test makes it sensitive to
differences in mortality in different directions at different ages.
They discussed the desirability of this feature and noted
that it should be possible to carry out the maximisation subject to the
weights being positive but ``could not see how'' to do this (the
derivation of a test designed to detect differences in the same
direction in all age groups was not to be proposed until the work of
Cochran, nearly forty years later---see our discussion of the
$2\times2\times K$ below).
However, they argued that the sensitivity of their test to
differences in death rates in different directions in different age
groups in fact represented an improvement over the comparison of
corrected, or standardized, rates since ``that idea is essentially
imperfect and does not really distinguish between differences in the
manner of dying''.

\subsection*{Further Application of the Method of Expected Numbers of Deaths}

As described in Section~2, \citet{Westergaard-1882} from the very
beginning emphasised
that expected numbers of death could be calculated
according to any
stratification, not just age. Encouraged by \citeauthor{Westergaard-1916}'s
(\citeyear{Westergaard-1916})
survey in English,
\citet{Woodbury-1922} demonstrated this through the example of infant
mortality as related
to mother's age, parity (called here \textit{order of birth}), earnings
of father
and plural births. For example, the crude death rates by order of
births form a
clear J-shaped pattern with nadir at third birth; assuming that only
age of the
mother was a determinant, one can calculate the expected rates for each
order of
birth, and one gets still a J, though somewhat attenuated, showing
that a bit of
the effect of birth order is explained by mother's age. Woodbury did
not forget
to warn:

\begin{quote}
Since it is an averaging process the method will yield satisfactory
results only
when an average is appropriate.
\end{quote}

\noindent\citet{Stouffer-Tibbitts-1933} followed up by pointing out
that in many
situations
the calculations of expected numbers for $\chi^2$ tests would coincide
with the ``Westergaard method''.

\section{Standardization in the 20{th}
Century}
\label{sec:stand20}
Although, as we have seen, standardisation methods were widely used in
the 19{th} century, it was in the 20{th} century
that a more careful examination of the properties of these methods was
made. Particularly important are the authoritative reviews by
\citet{Yule-1934} and, thirty years later, by Kitagawa
(\citeauthor{Kitagawa-1964}, \citeyear{Kitagawa-1964,Kitagawa-1966}).
Both these authors saw the
primary aim as being the construction of what Yule termed ``an average
ratio of mortalities'', although Yule went on to remark:

\begin{quote}
in Annual Reports and Statistical Reviews the
process is always carried a stage further, viz. to the calculation of
a ``standardized death-rate''. This extension is really superfluous,
though it may have its conveniences
\end{quote}

\noindent(the standardized rate in the study population being
constructed by
multiplying the crude rate in
the standard population by the standardized ratio of rates for the
study population versus the standard population).
\subsection*{Ratio of Averages or Average of Ratios?}
Both Yule and Kitagawa noted that central to the discussion was the
consideration of two sorts of indices. The first of these, termed
a ``ratio of averages'' by Yule, has the form
$\sum w_ix_i/\sum w_iy_i$, while the second, which he termed an
``average of ratios'', has the form $\sum w_i^*(x_i/y_i)/\sum w_i^*$.
Kitagawa noted that economists would describe the former as an
``aggregative index'' and the latter as an ``average of relatives''.

Both authors pointed out that, although the two types of
indexes seem to be doing rather different things, it is somewhat
puzzling that they are algebraically
equivalent---we only have to write $w_i^* = w_iy_i$.
It is important to note, however, that the algebraic equivalence does
not mean that a given index is equally interpretable in either
sense. Thus, for the index to be interpretable as a ratio of averages,
the weights $w_i$ must reflect some population distribution so that
numerator and denominator of the index represent marginal expectations
in the same population. Alternatively, to present the average of
the age-specific ratios, $x_i/y_i$,
as a single measure of the age-specific effect
would be misleading if they were not reasonably homogeneous.
Kitagawa concluded:

\begin{quote}
the choice between an aggregative index and an average of
relatives in a mortality
analysis, for example, should be made on the basis of whether the
researcher wants
to compare two schedules of death rates in terms of the total number of deaths
they would yield in a standard population \textit{or }in terms of the relative
(proportionate) differences between corresponding specific rates in
the two schedules.
Both types of index can be useful when correctly applied and interpreted.
\end{quote}

Here Kitagawa very clearly defined the distinction between what we, in
the \hyperref[sec:intro]{Introduction}, termed the \textit{marginal} and the \textit{conditional}
targets. Immediately after this definition, she hastened to point out
that:

\begin{quote}
It must be recognized at the outset, however, that no single summary statistic
can be a substitute for a detailed comparison of the specific rates in
two or more
schedules of rates.
\end{quote}

On the matter of averaging different ratios, \citet{Yule-1934} started his
paper with the example of comparing the death
rates for England and Wales for 1901 and 1931. His Table~I contains
these for both sexes in 5-year age groups and he commented:

\begin{quote}
\ldots the rates have fallen at all ages up to 75 for males and 85
for females. At the same time the amount of the fall is very
different at different ages, apart even from the actual rise in old
age. The problem is simply to obtain some satisfactory form of
average of all the ratios shown in columns 4 and 7, an average which
will measure in summary form the general fall in mortality between
the two epochs, just as an index-number measures the general fall or
rise in prices.
\end{quote}

So far, there is no requirement for these ratios to be
similar. However, when describing indirect standardisation,
\citeauthor{Yule-1934} [(\citeyear{Yule-1934}), page 12] pointed out that

\begin{quote}
if \ldots all the ratios of sub-rates are the same, no variation of
weighting can make any difference,
\end{quote}

\noindent and warned (page 13),

\begin{quote}
and perhaps it may be remarked that \ldots if the
ratios $m_{ur}/m_{sr}$ are very different in different age groups, any
comparative mortality figure becomes of questionable value.
\end{quote}

The issue of constancy of ratios was picked up in the printed
discussion of the paper [\citep{Yule-1934}, page 76] by Percy Stokes, seconder
of vote of thanks:

\begin{quote}
Those of us who have taught these methods to students have been
accustomed to point out that they lead to identical results when the
local rates bear to the standard rates the same proportion at every
age.
\end{quote}

\subsection*{Comparability of Mortality Ratios}
Yule noted that, particularly in official mortality statistics,
standardisation is applied to many different study populations so
that, as well as the standardized ratio of mortality in each study
population to the standard population being meaningful in its own
right, the comparison of the indices for two study populations should
also be meaningful. He drew attention to the fact that the ratio of
two seemingly legitimate indices is not necessarily itself a legitimate
index. He
concluded that either type of index could legitimately be used
either if the same weights $w_i$ are used across study populations
(for ratios of averages) or if the same $w_i^*$ are used (for averages
of ratios).

\newcommand{\ppsub}[3]{{}_{#1}{#2}^{}_{#3}}
Denoting a standardized ratio for comparing study groups A and B with
standard by $\ppsub{s}{R}{a}$ and $\ppsub{s}{R}{b}$, respectively,
Yule suggested that
$\ppsub{s}{R}{a}/\ppsub{s}{R}{b}$ should be a legitimate index of the
ratio of mortalities
in population A to that in population B. He also
suggested that, ideally, $\ppsub{a}{R}{b} =
\ppsub{s}{R}{a}/\ppsub{s}{R}{b}$
but noted that, whereas the CMF of
direct standardisation fulfills the former criterion, no method of
standardisation hitherto suggested fulfilled this more stringent
criterion. Indirect standardisation fulfils neither criterion and Yule
judged it to be ``hardly a method of standardisation at all''.

Yule's paper is also famous for its derivation of standard errors of
comparative mortality figures; for the particular case of the SMR, we have
\[
\operatorname{SMR} = \mbox{Observed}/\mbox{Expected},\quad O/E
\]
and
\[
\mbox{S.E.(SMR)}\approx\sqrt{O}/E.
\]
As noted earlier, this was already derived by Westergaard (\citeyear{Westergaard-1882}),
although this was apparently not generally known.

A final matter occupying no less than twelve pages of \citet{Yule-1934} is
the discussion of a context-free average, termed by Yule his $C_3$ method
or the \textit{equivalent average death rate}, which is just the simple average
of all age-specific death rates. This quantity could also be explained
as the death rate standardized to a population with equal numbers in
each age group. As we shall see below, it was further discussed by
\citet{Kilpatrick-1962} and rediscovered by \citet{Day-1976}
in an application to
cancer epidemiology. In modern survival analysis it is called the
cumulative hazard and estimated nonparametrically by the Nelson--Aalen
estimator (\citep{nelson:72}, \citep{aalen:78}, \citep{andeborg}).

\subsection*{Elaboration: Rosenberg's Test Factor Standardisation}

During World War II, the United States Army established a Research
Branch to investigate problems of morale, soldier preferences and
other issues to provide information that would allow the military to
make sensible decisions on practical issues involved in army life. To
formalize some of the tools used in that generally rather practical
research, illustrated with concrete examples from that work,
\citet{Kendall-Lazarsfeld-1950}
introduced and discussed the terminology of
elaboration: A statistical relation has been established between two
variables, one of which is assumed to be the cause, the other to be
the effect. The aim is to further understand that relation by introducing
a third variable (called test factor) related to the ``cause'' as well
as the ``effect''. Kendall and Lazarsfeld carefully distinguished
between antecedent and intervening test variables, depending on the
temporal order of the ``cause'' and the test variables. If the
population is stratified according to an antecedent test factor, and
the partial relationships between the two original variables then
vanish, the relation between ``cause'' and ``effect'' has been explained
through their relations to the test variable, which is then termed
spurious. If the association between cause and effect disappears (is
reduced) by controlling on the intervening variable, Kendall and
Lazarsfeld talk about complete (partial) interpretation of the
original two-factor relationship.

We note that interpretation has gone out of use at least in
epidemiological applications and in most of modern causal inference
where the focus is on obtaining an undiluted measure of the causal
effect of the ``cause'', not diluting this effect by conditioning on
variables on the causal pathway from cause to effect [see \citet{Pearl-2001}
or \citet{Petersen-etal.-2006}]. Instead, a general area of Mediation
Analysis has grown up; see \citet{MacKinnon-2008}, Section~1.8, for a useful
historical survey.

\citet{Rosenberg-1962} used standardization to obtain a single summary
measure from all the partial (i.e., conditional) associations resulting
from the stratification in an elaboration. Rosenberg's famous example
was a study of the possible association between religious affiliation
and self-esteem for high school students, controlling for (all
combinations of) father's education, social class identification and
high school grades. Thus, this is an example of interpretation by
conditioning on variables that might mediate an effect of religious
affiliation on sons' self-esteem. The crude association showed higher
self-esteem for Jews than for Catholics and Protestants; by
standardizing on the joint distribution of the three covariates in the
total population this difference was halved.

Rosenberg emphasized that in survey research the end product of the
standardisation exercise is not a single rate as in demography, but:

\begin{quote}
In survey research, however, we are interested in \textit{total
distributions}. Thus, if we examine the association between $X$ and $Y$
standardizing on $Z$, we must emerge with a standardised table (of the
joint distribution of $X$ and $Y$) which contains all the cells of the
original table.
\end{quote}

Rosenberg indicated shortcuts to avoid repeating the same calculations
when calculating the entries of this table.

\subsection*{The Peters--Belson Approach}
This technique (\cite{Peters-1941,Belson-1956})
was developed for comparing an experimental group with a control
group in an observational study on some continuous outcome. The
proposal is to
regress the outcome on covariates only in the control group and use
the resulting
regression equation to predict the results for the experimental group
under the
assumption of no difference between the groups. A simple test of no differences
concludes the analysis. \citet{Cochran-1969} showed that under some
assumptions of
(much) larger
variance in the experimental group than the control group this
technique might
yield stronger
inference than standard analysis of covariance, and that it will also
be robust
to certain types of effect modification. The technique has recently
been revived
by \citet{Graubard-etal.-2005}.

\subsection*{Decomposition of Crude Rate Differences and Ratios}
\label{sec:decomp}

Several authors have suggested a decomposition of a contrast between
two crude rates into a component due to differences between the
age-specific rates and a component due to differences between the age
structures of the two populations.

\citet{Kitagawa-1955} proposed an additive decomposition in which the
difference in crude rates is expressed as a sum of (a) the difference
between the (direct) standardized rates, and (b) a residual due to the
difference in age structure. Rather than treating one population as the
standard population and the second as the study population, she
treated them symmetrically, standardising both to the mean of the two
populations' age structures:
\begin{eqnarray*}
&& \mbox{Crude rate (study)} - \mbox{Crude rate (standard)}
\\
&&\quad= \sum a_i\alpha_i - \sum
s_i\lambda_i
\\
&&\quad= \sum(\alpha_i - \lambda_i)
\frac{a_i + s_i}{2} + \sum(a_i - s_i)
\frac{\alpha_i + \lambda_i}{2}.
\end{eqnarray*}
The first term contrasts the standardized rates while the second
contrasts the age structures.

However, ratio comparisons are more frequently employed when
contrasting rates and several authors have considered a
multiplicative decomposition in which the ratio of crude rates is
expressed as the product of a standardized rate ratio and a factor
reflecting the effect of the different age structures. Such a
decomposition, in which the age-standardized measure is the SMR, was
proposed by \citet{Miettinen-1972}:
\begin{eqnarray*}
\frac{\mbox{Crude rate (study)}}{\mbox{Crude rate (standard)}} &=& \frac
{\sum a_i \alpha_i}{\sum s_i \lambda_i}
\\
&=& \frac{\sum a_i \alpha_i}{\sum
a_i \lambda_i} \times
\frac{\sum a_i \lambda_i}{\sum s_i \lambda_i}.
\end{eqnarray*}
The first term is the SMR and the second, which reflects the effect of
the differing age structures, Miettinen termed the ``confounding risk
ratio''.

\citet{Kitagawa-1964} had also
proposed a multiplicative decomposition which, as in her
additive decomposition, treated the two populations
symmetrically. Here, the standardized ratio measure was inspired by
the literature on price indices in economics. If, in a ``base''
year, the price of commodity $i$ is $p_{0i}$ and the quantity
purchased is $q_{0i}$ and, in year $t$ the equivalent values are
$p_{ti}$ and $q_{ti}$, then an overall comparison of prices requires
adjustment for differing consumption patterns. Simple relative indices
can be constructed
by fixing consumption at base or at $t$. The former is
Laspeyres's index, $\sum p_{ti}q_{0i}/\sum p_{0i}q_{0i}$, and the
latter is Paasche's index, $\sum p_{i}q_{ti}/\sum p_{0i}q_{ti}$.
These are asymmetric with respect to the
two time points and this asymmetry is addressed in Fisher's ``ideal''
index, defined as the geometric mean of Laspeyres's and Paasche's
indices. Kitagawa noted that Laspeyres's and Paasche's indices are
directly analogous to the CMF and SMR, respectively, and, in her symmetric
decomposition,
\[
\frac{\sum a_i\alpha_i}{\sum s_i\lambda_i} = \sqrt{ \frac{\sum s_i\alpha
_i}{\sum s_i\lambda_i}\times\frac{\sum a_i\alpha_i}{\sum a_i\lambda_i}
} \times
\sqrt{ \frac{\sum\lambda_ia_i}{\sum\lambda_is_i}\times\frac{\sum
\alpha_ia_i}{\sum\alpha_is_i} },
\]
the first term is an ``ideal'' index
formed by the geometric mean of the CMF and SMR, and the second term is:

\begin{quote}
the geometric mean of two indexes summarizing differences in
$I$-composition; one an aggregative index using the $I$-specific rates
of the base population
as weights, and the second an aggregative index using the $I$-specific rates
of the given population as weights.
\end{quote}

The paper by \citet{Kitagawa-1955} concluded with a detailed comparison to the
``Westergaard method'' as documented by \citet{Woodbury-1922}. Woodbury's
paper had also inspired Kitagawa's contemporary R. H. Turner, also a
Ph.D. from the University of Chicago, to develop an approach to
additive decomposition according to several covariates (\citep{Turner-1949}),
showing how the ``nonwhite--white'' differential in labour force
participation is associated with marital status, household
relationship and age. Kitagawa's decomposition paper continues to be
frequently cited and the technique is still included in current
textbooks in demography [e.g., \citet{Preston-etal.-2001}]. There has
been a
considerable further development of additive decomposition ideas; for
recent reviews see \citet{Chevan-Sutherland-2009} for the development in
demography and \citet{Powers-Yun-2009} for decomposition of hazard rate
models and some references to developments in econometrics and to some
extent in sociology. We return in Section~\ref{sec:models}
to the connection with the
method of ``purging'' suggested by C. C. Clogg.

\section{Odds Ratios and the \texorpdfstring{$2\times2\times K$}{$2times2timesK$} Contingency Table}
\label{sec:cc}
\subsection*{Case--Control Studies and the Odds Ratio}
Although the case-control study has a long history, its use to provide
quantitative measures of the strength of association is more recent,
generally being attributed to \citet{cornfield:1951}. Table~\ref{tab:cc22}
sets out results from a hypothetical
case--control study comparing some exposure in cases of a disease with
that in a control group of individuals free of the disease.
%
\begin{table}
\caption{Frequencies in a $2\times2$ contingency table derived
from a case--control study}
\label{tab:cc22}
\begin{tabular}{cc}
\begin{tabular}{@{}lcc@{}}
\hline
& \textbf{Cases} & \textbf{Controls}\\
\hline
Exposed & $A$ & $B$\\
Not exposed & $C$ & $D$\\
\hline
\end{tabular}
&
$N = A + B + C +D$
\end{tabular}
\end{table}
In this work, he
demonstrated that, if the disease is rare, that is,
prevalence of disease in the population, $X$, is near zero
and the proportion of cases and controls exposed are $p_1$
and $p_0$, respectively, then the prevalence of disease in exposed
subjects is, to a close approximation, $X p_1/p_0$, and $X
(1-p_1)/(1-p_0)$ in subjects not exposed. Thus, the ratio of
prevalences is approximated by the \textit{odds ratio}
\[
\frac{p_1}{1-p_1} \Big/\frac{p_0}{1-p_0},
\]
which can be estimated by $(AD)/(BC)$.

In this work, Cornfield discussed the problem of bias due to poor
control selection, but did not explicitly address the problem of
confounding by a third factor.
In later work Cornfield (\citeyear{Cornfield-1956}) did consider the case
of the $2\times2 \times K$ table in which the $K$ strata were
different case--control studies. However, his analysis focussed on the
\textit{consistency} of the stratum-specific odds ratios; having
excluded outlying studies, he, at this stage,
ignored Yule's paradox,
simply summing over the remaining studies and calculating the odds
ratio in the marginal $2\times2$ table.
\subsection*{Interaction and ``Simpson's Paradox''}
Bartlett (\citeyear{Bartlett-1935}) linked consistency of odds ratios in
contingency tables with the concept of ``interaction''. Specifically,
he defined zero second order interaction in the $2\times2\times2$
contingency table of variables $X$, $Y$ and $Z$ as occurring when the
odds ratios between $X$ and $Y$ conditional upon the level of $Z$ are
stable across levels of $Z$. (Because of the symmetry of the odds
ratio measure, the roles of the three variables are
interchangeable.) In an important and much cited paper,
\citet{Simpson-1951} discussed interpretation of no interaction in the
$2\times2\times2$ table, noting that ``there is considerable scope
for paradox''.

If one were to read only the abstract of Simpson's paper, one could be
forgiven for believing that he had simply restated Yule's paradox in
this rather special case:

\begin{quote}
it is shown by an example that vanishing of this second order
interaction does not necessarily justify the mechanical procedure of
forming the three component $2\times2$ tables and testing each of
these for significance by standard methods
\end{quote}

\noindent(by ``component'' tables, he meant the marginal tables). Thus,
``Simpson's paradox'' is often identified with Yule's paradox,
sometimes being referred to as the Yule--Simpson paradox. However, the
body of Simpson's paper contains a much more subtle point about the
nature of confounding.

Simpson's example is a table in which $X$ and $Y$ are both associated
with $Z$, in which there is no second order interaction, and the
conditional odds ratios for $X$ versus $Y$ are 1.2 while the marginal
odds ratio is 1.0. He pointed out that if $X$ is a medical treatment,
$Y$ an outcome and $Z$ sex, then there is clearly a treatment effect---the
conditional odds ratio provides the ``right'' answer, the
treatment effect having been destroyed in the margin by negative
confounding by sex. Simpson
compared this with an imaginary experiment concerning a pack of
playing cards which have been played with by a baby in such a way that
red cards and court cards, being more attractive, have become
dirtier. Variables $X$ and $Y$ now denote red/black and court/plain
and $Z$ denotes the cleanliness of the cards. In this case, Simpson
pointed out that the \textit{marginal} table of $X$ versus $Y$,
``provides what
we would call the sensible answer, that there is no such
association''. This is, perhaps, the real Simpson's paradox---the same
table demonstrates Yule's paradox when
labelled one way but does not when it is labelled another
way. Simpson's paper pointed out that the causal status of variables
is central; one can condition on \textit{causes} when forming conditional
estimates of treatment effects, but not upon \emph{effects}. As we
shall see in the next section, this point is central
to the problem of
time-dependent confounding which has inspired much
recent methodological advance. A closely related issue is the
phenomenon of selection bias, famously discussed by \citet
{Berkson-1946} in
relation to hospital-based studies. There $X$ and $Y$ are observed only
when an effect, $Z$ (e.g., attending hospital), takes on a specific
value.

A further contribution of Simpson's paper was to point out the
``noncollapsibility'' of the odds ratio measure in this zero
interaction case; the conditional and marginal odds ratios between $X$
and $Y$ are only the same if either $X$ is conditionally independent
of $Z$ given $Y$, or $Y$ is conditionally independent of $Z$ given~$X$.
Note that these conditions may not be satisfied even in
randomised studies---another of the paradoxes to which Simpson drew
attention. For a more detailed discussion of Simpson's paper see
\citet{Hernan-Clayton-Keiding-2011}.
\subsection*{Cochran's Analyses of the $2\times2\times K$ Table}
In his important paper on ``methods for strengthening the common
$\chi^2$ test'', \citet{cochran:54} proposed a ``combined test of
significance of the difference in occurrence rates in the two
samples'' when ``the whole procedure is repeated a
number of times under somewhat differing environmental conditions''.
He pointed out that carrying out
the $\chi^2$ test in the marginal table

\begin{quote}
is legitimate only if the probability $p$ of an
occurrence (on the null hypothesis) can be assumed to be the same in
all the individual $2\times2$ tables.
\end{quote}

\noindent (He did not further qualify this statement in the light of Simpson's
insight discussed above.)
He proposed three alternative analyses.
The first of these was to add up the $\chi^2$ test statistics from
each table and to compare the result with the $\chi^2$ distribution
on $K$ degrees of freedom. This, as already noted, is equivalent to
Pearson and Tocher's earlier proposal, but Cochran judged it a poor
method since

\begin{quote}
It takes no account of the signs of the differences
$(p_1 - p_0)$ in the two samples, and consequently lacks power in
detecting a difference that shows up consistently in the same
direction in all or most of the individual tables.
\end{quote}

The second alternative he considered was to calculate the ``$\chi$''
value for each table---the square roots of the $\chi^2$ statistics,
with signs equal to those of the corresponding $(p_1 - p_0)$'s---and
to compare the sum of these values
with the normal distribution with mean zero and
variance $K$. He noted, however, that this method would not be
appropriate if the sample sizes (the ``$N's$'') vary
substantially between tables, since

\begin{quote}
Tables that have very small $N$'s cannot be
expected to be of much use in detecting a difference, yet they receive
the same weight as tables with large $N's$.
\end{quote}

He also noted that variation of the probabilities of outcome between
tables would also adversely affect the power of this method:

\begin{quote}
Further, if the $p$'s vary from say 0 to 50\%, the difference that
we are trying to detect, if present, is unlikely to be constant at
all levels of $p$. A large amount of experience suggests that the
difference is more likely to be constant on the probit or logit scale.
\end{quote}

It is clear, therefore, that Cochran considered the ideal analysis to
be based on a model of ``constant effect'' across the tables. Indeed,
when the data were sufficiently extensive, he advocated use of
empirical logit or probit transformation of the observed proportions
followed by model fitting by weighted least squares. Such an approach, based
on fitting a formal model to a table of proportions, had already been
pioneered by \citet{dyke:patterson:52}, and will be discussed in
Section~\ref{sec:models}.

In situations in which the data were not sufficiently extensive to
allow an approach based on empirical transforms, Cochran proposed an
alternative test ``in the original scale''. This involved
calculating a weighted mean of the differences $d = (p_1 - p_0)$ over
tables. In our notation, comparing the prevalence of exposure between
cases and controls,
\begin{eqnarray*}
d_i &=& \frac{A_i}{A_i+C_i} - \frac{B_i}{B_i+D_i},
\\
w_i &=& \biggl(\frac{1}{A_i+C_i} + \frac{1}{B_i+D_i}
\biggr)^{-1},
\\
\overline{d} &=& \sum w_id_i \big/ \sum
w_i.
\end{eqnarray*}
In calculating the variance of $\overline{d}$, he estimated the
variance of the $d_i$'s under a binomial model using a plug-in
estimate for the expected values of $p_{1i}, p_{0i}$ under the null
hypothesis: $(A_i+B_i)/N_i$. Cochran described the resulting test as
performing well ``under a wide range of variations in the $N$'s and
$p$'s from table to table''.

A point of some interest is Cochran's choice of weights which, as
pointed out by \citet{Birch-1964}, was ``rather heuristic''. If this
procedure had truly been, as Cochran described it, an analysis ``in the
original scale'', one would naturally have weighted the differences
inversely by their variance. But this does not lead to Cochran's
weights, and he provided no justification for his alternative
choice. A likely possibility is that he noted that weighting
inversely by precision leads to two different tests according to
whether we choose to compare the proportions exposed between cases and
control or the proportions of cases between exposed and unexposed
groups. Cochran's choice of weights avoided this embarrassment.
\subsection*{Mantel and Haenszel}
Seemingly unaware of Cochran's work, \citet{mantel:haenszel:1959}
considered the analysis of the $2\times2\times K$ contingency table.
This paper explicitly related the discussion to control for
confounding in case--control studies. Before discussing this famous
paper, however, it is interesting that the same authors had suggested
an alternative approach a year earlier
(\citep{haenszel:shimkin:mantel:1958}).

As in Cochran's analysis, the idea was based on
post-stratification of cases and controls into strata which are as
homogeneous as possible. Arguing by analogy with the method of indirect
standardisation of rates, they suggested that the influence of
confounding on the odds ratio could be assessed by calculating, for
each stratum, $s$, the ``expected'' frequencies in the $2\times2$
table under the assumption of no partial association within
strata and calculating the marginal odds ratio under this
assumption. The observed marginal odds ratio was then adjusted by this
factor.
Thus, denoting the expected frequencies by $a_i, b_i, c_i$ and $d_i$ where
$a_i = (A_i+B_i)(A_i+C_i)/N_i$ etc., their proposed index was
\[
\frac{\sum A_i \sum D_i}{\sum B_i \sum C_i} \Big/ \frac{\sum a_i \sum
d_i}{\sum b_i \sum c_i}.
\]
The use of the stratum-specific expected frequencies in this way can
be regarded as an early attempt, in the case--control setting,
to estimate what later became known as
the ``confounding risk ratio'' and which we described in
Section~\ref{sec:decomp}.

In their later paper, \citet{mantel:haenszel:1959}
themselves criticized this adjusted index which, they stated,
``can be seen to have a bias toward unity'' and does ``not yield an
appropriate adjusted relative risk''. (Somewhat unconvincingly,
they claimed that they
had used the index fully realizing its deficiencies ``to present
results more nearly comparable with those reported by other
investigators using similarly biased estimators''!)
These statements were not
formally justified and beg the question as to what, precisely, is the
estimand? One can only assume that they were referring to the case in
which the stratum-specific odds ratios
are equal and provide a single estimand. This is the case in which
Yule's $Q$ is stable across subgroups. The alternative estimator they
proposed:
\[
\frac{\sum A_i D_i/N_i}{\sum B_iC_i/N_i}
\]
is a consistent estimator of the stratum-specific odds ratio in this
circumstance. They also proposed a test for association between
exposure and disease within strata. The test statistic is the sum,
across strata, of
the differences between observed and ``expected'' frequencies in one
cell of each table:
\begin{eqnarray*}
\sum(A_i - a_i) &=& \sum
A_i - \frac{(A_i+B_i)(A_i+C_i)}{N_i}
\\
&=& \sum\frac{1}{N_i} (A_iD_i -
B_iC_i ),
\end{eqnarray*}
and its variance under the null hypothesis is
\[
\sum\frac{(A_i+B_i)(C_i+D_i)(A_i+C_i)(B_i+D_i)}{N_i^2(N_i-1)}.
\]
Some algebra shows that the Mantel--Haenszel test statistic is
identical to Cochran's $\sum w_id_i$. There is a slight difference
between the two procedures in that, in calculating the variance,
Mantel and Haenszel used a
hypergeometric assumption to avoid the need to estimate a nuisance
parameter in each stratum in the ``two binomials'' formulation. This
results in the $(N_i-1)$ term in the above variance formula instead of
$N_i$---a~distinction which can become important when there are a
large number of sparsely populated strata.

Whereas considerations of bias and, as later shown, optimal properties
of their proposed test depend on the assumption of constancy of the
odds ratio across strata, Mantel and Haenszel were at pains to disown
such a model. They proposed that any standardized, or corrected,
summary odds ratio would be some sort of weighted average of the
stratum-specific odds ratios and identified that one might choose
weights either by \textit{precision} or by \textit{importance}. On the
former:

\begin{quote}
If one could assume that the increased relative risk associated
with a factor was constant over all subclassifications, the
estimation problem would reduce to weighting the several
subclassification estimates according to their relative
precisions. The complex maximum likelihood iterative procedure
necessary for obtaining such a weighted estimate would seem to be
unjustified, since the assumption of a constant relative risk can
be discarded as usually untenable.
\end{quote}

They described the weighting scheme used in the Mantel--Haenszel
estimator as approximately weighting by precision. Indeed, it turns
out that these weights correspond to optimal weighting by precision
for odds ratios close to 1.0.

An alternative standardized odds ratio
estimate, in the spirit of weighting and mirroring direct
standardisation, was proposed by \citet{miettinen:1972a}. This is
\[
\frac{\sum W_i A_i/B_i}{\sum W_i C_i/D_i},
\]
where the weights reflect the population distribution of the
stratifying variable. This index can be unstable when strata are
sparse, but \citet{Greenland-1982} pointed out that it has clear
advantages over the Mantel--Haenszel estimate when the odds ratios
differ between strata. This follows from our earlier discussion
(Section~\ref{sec:stand20}) of the
distinction between a ratio of averages and an average of
ratios. Since the numerator and denominator of the Mantel--Haenszel
estimator do not have an interpretation in terms of the population
average of a meaningful quantity, the index must be interpreted as an
average of ratios, despite its usual algebraic representation. Thus,
despite the protestations of Mantel and Haenszel to the contrary, its
usefulness depends on approximate stability of the stratum-specific
odds ratios. Greenland pointed out that Miettinen's index has an
interpretation as a ratio of marginal expectations of
epidemiologically meaningful quantities and, therefore, may be useful
even when odds ratios are heterogeneous. He went on to propose some
improvements to address its instability.

As was noted earlier, there was a widespread belief that controlling
for confounding in case-control studies was largely a matter to be
dealt with at the design stage, by appropriate ``cross-matching'' of
controls to cases. Mantel and Haenszel, however, pointed out that such
matching nevertheless needed to be taken account of in the analysis:

\begin{quote}
when matching is made on a large number of factors, not
even the fiction of a random sampling of control individuals can be
maintained.
\end{quote}

They showed that the test and estimate they had proposed were still
correct in the setting of closely matched studies. Despite this,
misconceptions about matching persisted for more than a decade.

\section{The Emergence of Formal Models}
\label{sec:models}

Except for linear regression analysis for quantitative data, proper statistical
models, in the sense we know today, were slow to appear for the
purpose of what we now call confounder control.

We begin this section with the early multiplicative intensity
age-cohort model
for death rates by
\citeauthor{Kermack-McKendrick-McKinlay-1934a}
(\citeyear{Kermack-McKendrick-McKinlay-1934a,Kermack-McKendrick-McKinlay-1934b}),
even though it was
strangely isolated
as a statistical innovation: no one outside of a narrow circle of
cohort analysts
seems to have quoted it before 1976. First, we must mention
two precursors from the
actuarial environment.

\subsection*{Actuarial Analyses of Cohort Life Tables}
Two papers were read to audiences of actuaries on the same evening: 31 January
1927.
\citet{Derrick-1927}, in the Institute of Actuaries in London, studied
mortality in England
and Wales 1841--1925, omitting the war (and pandemic) years 1915--1920.
On a clever
graph of age-specific mortality (on a logarithmic scale) against year
of birth
he generalized the parallelism of these curves to a hypothesis that
mortality was
given by a constant age structure, a decreasing multiplicative
generation effect
and no period effect, and even ventured to extrapolate the mortality
for existing
cohorts into the future.

Davidson and Reid, in the Faculty of Actuaries in Edinburgh, first gave
an exposition of estimating mortality rates in a Bayesian framework (posterior
mode), including the maximum likelihood estimator interpretation of
the empirical
mortality obtained from an uninformative
prior (\citep{Davidson-Reid-1926-1927}). They proceeded to
discuss how the
\textit{mortality variation force} might possibly depend on age and
calendar year
and arrived at a discussion on how to predict future mortality, where
they remarked (page 195) that this would be much easier if

\begin{quote}
there is in existence a law of mortality which, when applied to {\it
consecutive}
human life---that is, when applied to trace individuals born in a
particular calendar
year throughout the rest of their lives---gives satisfactory results
\end{quote}

\noindent or, as we would say, if the cohort life table could be
modelled. Davidson and Reid also
explained their idea through a well-chosen, though purely theoretical, graph.

\subsection*{The Multiplicative Model of Kermack,
McKendrick and McKinley}

Kermack, McKendrick and McKinley published an analysis of death-rates
in England
and Wales since 1845, in Scotland since 1860 and in Sweden since 1751
in two companion
papers.
In the substantive presentation in The Lancet
(Kermack, Mc{K}endrick and Mc{K}inlay, \citeyear{Kermack-McKendrick-McKinlay-1934a})---republished by International
Journal of Epidemiology (2001) with discussion of the epidemiological
cohort analysis
aspects---they observed and discussed a clear pattern in these rates
as a product
of a factor only depending on age and a factor only depending on year
of birth.

The technical elaboration in the Journal of Hygiene
(\citep{Kermack-McKendrick-McKinlay-1934b}) started from the partial
differential equation describing age-time dependent population growth
with $\nu_{t,a} da$ denoting the number of persons at time
$t$ with age between $a$ and $a + da$, giving the
death rate at time $t$ and age $a$
\[
-\frac{1}{\nu_{t,a}} \biggl( \frac{\partial\nu_{t,a}}{\partial t} +
\frac{\partial\nu_{t,a}}{\partial a} \biggr) =
f(t, a),
\]
here quoted from \citet{McKendrick-1925-1926}
[cf.  \citet{Keiding-2011} for comments on the history
of this equation], and postulate at once the multiplicative model for
\[
f(t, a) = \alpha(t - a)\beta_{a}.
\]
The paper is largely concerned with estimation of the parameters and
of the standard
errors of these estimates; some attention is also given to the
possibility of fitting
the age effect $\beta_{a}$ to the Gompertz--Makeham distribution.

This fine statistical paper was quoted very little in the following 45
years and
thus does not seem to have influenced the further developments of
statistical models
in the area. One cannot avoid speculating what would have happened if
this paper
had appeared in a statistical journal rather than in the Journal of
Hygiene. 1934
was the year when Yule had his major discussion paper on
standardisation in the
Royal Statistical Society. In all fairness, it should, on the other
hand, be emphasised
that Kermack et al. did not connect to the then current discussions of general
issues of standardisation.

\subsection*{The SMR as Maximum Likelihood Estimator}

Kilpatrick (\citeyear{Kilpatrick-1962}), in a paper based on his Ph.D.
at Queen's
University at Belfast,
specified for the first time a mortality index as an estimator of
a parameter in a well-specified statistical model---that in which
the age-specific relative death rate in each age group estimates a
constant, and only differs from
it by random variation.
Kilpatrick's introduction is a good example of a statistical view on
standardization, in some ways rather reminiscent of Westergaard:

\begin{quote}
The mortality experienced by different  groups of individuals is best
compared, using specific death rates of sub-groups alike in every
respect, apart from the single factor by which the total population is
divided. This situation is rarely, if ever, realized and we have to be
satisfied with mortality comparisons between groups of individuals
alike with regard to two, three or four major factors known to affect
the risk of death.

In this paper groups are defined as aggregates of occupations (social
classes). It is assumed that age is the only factor related to an
individual's mortality within a group. This example may readily be
extended to other factors such as sex, marital status, residence,
etc. Although the association of social class and age-specific
mortality may be evaluated by comparisons between social classes,
specific death rates of a social class are more frequently compared
with the corresponding rates of the total population. It is this type
of comparison which is considered here.
\end{quote}

Kilpatrick then narrowed the focus to developing an index $I_{us}$
which

\begin{quote}
should represent the ``average'' excess or deficit mortality in group $u$
compared with the standard $s$,
\end{quote}

\noindent and noted that, with $\theta_x$ representing the ratio
between the
mortality rates in age group $x$ in the study group and the total population,

\begin{quote}
Recent authors \ldots have shown that the SMR
can be misleading if there is much variation in $\theta_x$ over the
age range considered. It would, therefore, seem desirable to test the
significance of this variation in $\theta_x$ before placing confidence on
the results of the SMR or any other index.
\ldots
This paper proposes a simple test for heterogeneity in $\theta_x$ and
shows that the SMR is equivalent to the maximum likelihood estimate of
a common $\theta$ when the $\theta_x$ do not differ significantly. It
follows therefore that the SMR has a minimum standard error.
\end{quote}

Formally, Kilpatrick assumed the observed age-specific rates in the
study group to follow Poisson distributions with rate parameters
$\theta\lambda_i$. The $\lambda_i$'s and the denominators, $A_i$,
were treated as deterministic constants, and the mortality ratio,
$\theta$, as a parameter to be estimated.

We note that the view of standardisation
as an estimation
problem in a well-specified statistical model was principally
different from earlier
authors. One could refer to the paper by
\citet{Kermack-McKendrick-McKinlay-1934b}
discussed above (which specifies a similar model),
but they did not explicitly see their model as being related to
standardisation;
their paper has been quoted rarely and it seems that Kilpatrick was
unaware of it.

Once standardisation is formulated as an estimation problem, the
obvious question
is to find an \textit{optimal} estimator, and Kilpatrick showed that the
standardized
mortality ratio (SMR)
\[
{\fontsize{9.8}{12.5}{\selectfont
\widehat{\theta} = \frac{\mbox{Observed number of deaths in the study
population }}{
\mbox{Expected number of deaths in the study population}}}}
\]
has minimum variance among all indices, and that it is the maximum
likelihood estimator
in the model specified by deterministic standard age-specific death
rates and a constant age-specific rate ratio.

Kilpatrick noted that while the SMR is, in a sense, optimal for
comparing one
study group to a standard, the weights change from one study group to
the next
so that it cannot be directly used for comparing several groups. As
we have seen, this point had been made often before, particularly
forcefully by \citet{Yule-1934}.
Kilpatrick compared
the SMR to the comparative mortality index (CMF) obtained from direct
standardization
and to Yule's index (the ratio of ``equivalent death rates'', that is, direct
standardization
using equally large age groups).
He concluded:

\begin{quote}
Where appropriate and possible, a test of heterogeneity on age-specific
mortality ratios should precede the use of an index. When there is insufficient
information
to conduct the test of heterogeneity, conclusions based solely on the index
value
may apply to none of the individuals studied. Caution is strongly urged
in the
interpretation of mortality indices.
\end{quote}

\subsection*{Kalton---Statistical View of Standardisation in Survey
Research}
Kalton (\citeyear{Kalton-1968}) surveyed, from a rather mainstream statistical
view, standardisation as a technique for estimating the contrast
between two groups and to test the hypothesis that this contrast
vanishes. Kalton emphasized that

\begin{quote}
\ldots if the estimate is to be meaningful, there must be virtually no
``interaction'' effect in the variable under study between the groups
and the control variable (i.e., there must be a constant difference
in the group means of the variable under study for all levels of the
control variable), but this requirement may be somewhat relaxed for
the significance test.
\end{quote}

This distinction implies that the optimal weights are not the same for
the estimation problem and the testing problem. Without commenting on
the causal structure of ``elaboration'' \citet{Kalton-1968}
also gave further insightful technical statistical comments to
Rosenberg's example (see
above) and the use of optimum weights for testing no effect of
religious group.

Kalton seems to have been unaware of Kilpatrick's paper six years
earlier, but took a similar mainstream statistical view of
standardisation: that presentation of a single summary measure of the
within-stratum effect of the study variable implies a model
of no interaction between stratum and study variable.

\subsection*{Indirect Standardisation without External Standard}

Kilpatrick had opened the way to a fully model-based analysis of rates
in lieu
of indirect standardisation, and authoritative surveys based on this
approach were indeed published by \citet{Holford-1980},
\citet{Hobcraft-etal.-1982},
\citet{Breslow-etal.-1983},
\citet{Borgan-1984} and \citet{Hoem-1987}.
Still, modified versions of the old technique of indirect
standardization remained part of the tool kit for many years.

An interesting example is the attempt by \citet{Mantel-Stark-1968}
to standardize
the incidence of mongolism for both birth order and maternal age. Standardized
for one of these factors, the incidence still increased as function of
the other,
but the authors felt it

\begin{quote}
desirable to obtain some simple descriptive statistics by which the
reader could
judge for himself what the data showed.
\ldots
What was required was that we determine simultaneously a set of
birth-order category
rates which when used as a standard set gave a set of
indirect-adjusted maternal-age
category rates which in turn, when used as a standard set, implied the original
set of birth-order category rates.
\end{quote}

The authors achieved that through an iterative procedure, which always
converged
to ``indirect, unconfounded'' adjusted rates, where the convergent
solutions varied
with the initial set of standard rates, although they all preserved
the \textit{ratios}
of the various birth-order category-adjusted rates and the ratios of
the various
maternal-age category-adjusted rates. \citet{Osborn-1975} and
\citet{Breslow-Day-1975} formulated
multiplicative models for the rates and used the same iterative
indirect standardisation
algorithm for the parameters. Generalizing
Kilpatrick's model to multiple study groups, the age-specific rate in
age group $i$ and study group $j$ is assumed to be $\theta_j
\lambda_i$. Treating $\lambda_i$'s as known, the $\theta_j$'s can be
estimated by SMRs; the $\theta_j$'s can then be treated as known and
the $\lambda_i$'s estimated by SMRs (although the indeterminacy
identified by Mantel and Stark must be resolved, e.g., by
normalization of one set of parameters). See
\citet{Holford-1980} for the
relation of this algorithm to iterative proportional fitting of
log-linear models in contingency tables. Neither
Mantel and Stark, Osborn, nor Breslow and Day cited Kilpatrick or
Kermack, McKendrick and McKinlay.

\subsection*{Logistic Models for Tables of Proportions}
We have seen that \citet{cochran:54} had suggested that analysis
of the comparison of two groups with respect to a binary response in
the presence
of a confounding factor (an analysis of a $2\times2\times K$
contingency table) could be approached by fitting formal models to
the $2\times K$ table of proportions, using a transformation such as
the logit or probit transformation. But such analyses, given
computational resources available at that time, were extremely
laborious. Cochran cited the pioneering work of
\citet{dyke:patterson:52} who developed a method for fitting the logit
regression model to fitted probabilities of response,
$\pi_{ijk\ldots}$, in a table:
\[
\log\frac{\pi_{ijk\ldots}}{1- \pi_{ijk\ldots}} = \mu+ \alpha_i + \beta
_j +
\gamma_k + \cdots
\]
by maximum likelihood, illustrating this technique with an
analysis estimating the
independent contributions of newspapers, radio, ``solid'' reading
and lectures upon
knowledge of cancer. Initially they applied an empirical logit
transformation to the observed proportions, $p_{ijk\ldots}$,
and fitted a linear model by weighted least squares. They then
developed an
algorithm to refine this solution to the true maximum likelihood, an
algorithm which was later generalized by \citet{nelder:wedderburn:72}
to the wider class of generalized linear models---the now familiar iteratively reweighted least squares (IRLS) algorithm.
Since, in their example, the initial fit to the empirical data
provided a good approximation to the maximum likelihood solution, only
one or two steps of the IRLS algorithm were necessary---perhaps
fortunate since the calculations were performed without recourse to a
computer.

Although, in its title, Dyke and Patterson referred to their paper as concerning
``factorial arrangements'', they explicitly
drew attention to its uses in dealing
with confounding in observational studies:

\begin{quote}
It is important to realise that with this type of data there are likely
to be a number of factors which may influence our estimate of the
effect of say, solid reading but which have not been taken into account.
The point does not arise in the case of well conducted experiments but
is common in survey work.
\end{quote}

\subsection*{Log-Linear Models and the National Halothane Study}
Systematic theoretical studies of multiple cross-classifications of
discrete data date back at least to \citet{Yule-1900}, quoted above. For
three-way tables, \citet{Bartlett-1935} discussed estimation and hypothesis
testing regarding the second-order interaction, forcefully followed up
by \citet{Birch-1963} in his study of maximum likelihood estimation in the
three-way table.

However, as will be exemplified below in the context of The National
Halothane Study, the real practical development in the analysis of
large contingency tables needed large computers for the necessary
calculations. This development largely happened around 1970 (with many
contributions from L. A. Goodman in addition to those already
mentioned), and the dominating method was straightforward maximum
likelihood.
Particularly influential were the dissertation by
\citet{Haberman-1974}, which also included important software,
and the authoritative monograph by \citet{Bishop-Fienberg-Holland-1975}.

\subsubsection*{The National Halothane Study}
Halothane is an
anaesthetic which around 1960 was suspected in the U.S. for causing
increased rates of hepatic necrosis, sometimes fatal. A subcommittee
under the U.S. National Academy of Sciences recommended that a large
cooperative study be performed, and this was started in July 1963. We
shall here focus on the study of ``surgical deaths'', that is, deaths during
the first 6 weeks after surgery. The study was based on retrospective
information from 34 participating medical centres, who reported all
surgical deaths during the four years 1959--1962 as well as provided
information on a random sample of about 38,000 from the total of about
856,000 operations at these centres during the four years. The study
was designed and analysed in a collaborative effort between leading
biostatisticians at Stanford University, Harvard University and
Princeton University/Bell Labs and the report (\citep
{Bunker-etal.-1969}) is
unusually rich in explicit discussions about how to handle the
adjustment problem with the many variables registered for the patients
and the corresponding ``thin'' cross-classifications. For a very
detailed and informative review, see \citet{Stone-1970}. The main
problem in the statistical analysis was
whether the different anaesthetics were associated with different death
rates, after adjusting for a range of possible confounders, as we
would say today.
In a still very readable introduction by B. W. Brown
et al. it was emphasized (page~185) that

\begin{quote}
the analysis of rates and
counts associated with many background variables is a recurring and
very awkward problem. \ldots It is appropriate to create new methods for
handling this nearly universal problem at just this time. High-speed
computers and experience with them have now developed to such a stage
that we can afford to execute extensive manipulations repeatedly on
large bodies of data with many control variables, whereas previously
such heavy arithmetic work was impossible. The presence of the large
sample from the National Halothane Study has encouraged the
investigation and development of flexible methods of adjusting for
several background variables. Although this adjustment problem is not
totally solved by the work in this Study, substantial advances have
been made and directions for further profitable research are clearly
marked.
\end{quote}

The authors here go on to emphasise that the need for
adjustment is not restricted to ``nonrandomised'' studies.

\begin{quote}
Pure or
complete randomization does not produce either equal or conveniently
proportional numbers of patients in each class; attempts at deep
post-stratification are  doomed to failure because for several
variables the number of possible strata quickly climbs beyond the
thousands. \ldots Insofar as we want rates for special groups, we need
some method of estimation that borrows strength from the general
pattern of the variables. Such a method is likely to be similar, at
least in spirit, to some of those that were developed and applied in
this Study. At some stage in nearly every large-scale, randomized
field study (a large, randomized prospective study of postoperative
deaths would be no exception), the question arises whether the
randomization has been executed according to plan. Inevitably,
adjustments are required to see what the effects of the possible
failure of the randomization might be. Again, the desired adjustments
would ordinarily be among the sorts that we discuss.
\end{quote}

The National Halothane Study has perhaps become particularly famous
among statisticians for the early multi-way contingency table analyses
done by Yvonne M. M. Bishop supervised by F. Mosteller. This approach
is here termed ``smoothed contingency-table analysis'', reflecting the
above-mentioned recognized need for the analysis to ``borrow strength
from the general pattern''. Bishop did her Ph.D. thesis in this area;
cf. the journal publications (\citeauthor{Bishop-1969},
\citeyear{Bishop-1969,Bishop-1971}) and
combined efforts with S. E. Fienberg and P. Holland in their very
influential monograph on
``Discrete Multivariate Analysis''
(\citep{Bishop-Fienberg-Holland-1975}).
But the various versions of data-analytic
(i.e., model-free) generalizations of standardisation are also of
interest, at least as showing how broadly these statisticians
struggled with their task: to adjust discrete data for many covariates
in the computer age.

The analysis began with classical standardization
techniques (L. Moses), which were soon overwhelmed by the difficulty
in adjusting for more than one variable at a time. Most of the
subsequent approaches use a rather special form of stratification
where the huge, sparse multidimensional contingency table generated by
cross-classification of covariates other than the primary exposure
variables (the anaesthetic agents) are aggregated to yield ``strata''
with homogeneous death rates, the agents subsequently compared by
standardizing across these strata.
Several detailed techniques were developed for this purpose by
J. W. Tukey and colleagues, elaborately documented in the report and
briefly quoted by \citeauthor{Tukey-1979} (\citeyear{Tukey-1979,Tukey-1991}); however, criticisms
were also
raised (\citeauthor{Stone-1970}, \citeyear{Stone-1970}; \citeauthor{Scott-1978}, \citeyear{Scott-1978}) and
the ideas do not seem to have
caught on.

\subsection*{Clogg's ``Purging'' of Contingency Tables}

Clifford Clogg was a Ph.D. student of Hauser, Goodman and Kitagawa at
the University
of Chicago, writing his dissertation in 1977 on Hauser's theme of
using a broader
measure of \textit{under}employment (as opposed to just \textit{un}employment)
as social indicator, in the climate of Goodman's massive recent
efforts on loglinear
modelling and Kitagawa's strong tradition in standardisation. We shall briefly
present Clogg's attempts at combining the latter two worlds in the
\textit{purging} techniques [\citet{Clogg-1978}, \citet{Clogg-Eliason-1988} and many
other articles]. A useful
concise summary was provided by \citeauthor{Sobel-1996}
[(\citeyear{Sobel-1996}), pages 11--14]
in his tribute to Clogg after Clogg's early death,
and a recent important discussion and
generalization was given by \citet{Yamaguchi-2011}.

Clogg considered a \textit{composition} variable $C$ with categories $i =
1,\ldots,I$, a \textit{group} variable $G$ with categories
$j=1,\ldots,J$, and a \textit{dependent} variable $D$ with categories
$k=1,\ldots,K$. The composition variable may itself have been generated
by cross-classification of several composition variables. The object
is to assess
the possible association of \textit{D }with \textit{G }adjusted for differences
in the compositions across the groups. Clogg assumed that the
three-way
$C\times G\times D$ classification
has already been modelled by a loglinear model, and the purging
technique was primarily
promoted as a tool for increased accessibility of the results of that analysis.
Most of the time the saturated model is assumed, although in our view
the purging
idea is much easier to assimilate when there is no three-factor interaction.

A brief version of Clogg's explanation is as follows. The $I\times
J\times K$
table is modelled by the saturated log-linear model
\[
\pi_{ijk} = \eta\tau^{\mathrm{ C}}_i
\tau^{\mathrm{ G}}_j \tau^{\mathrm{ D}}_k
\tau^{\mathrm{ CG}}_{ij} \tau^{\mathrm{ CD}}_{ik}
\tau^{\mathrm{ GD}}_{jk} \tau^{\mathrm{ CGD}}_{ijk}
\]
where the disturbing interaction is
$\tau^{\mathrm{ CG}}_{ij}$;
the  compo\-sition-specific rate
\[
r_{ij(k)} = \pi_{ijk} \Big/ \sum
_k \pi_{ijk} = \pi_{ijk}/
\pi_{ij\cdot}
\]
is independent of
$\tau^{\mathrm{ CG}}_{ij}$, but
the overall rate of occurrence
\[
r_{\cdot j(k)} = \sum_i\pi_{ijk}\Big /
\sum_{i,k} \pi_{ijk} =
\pi_{\cdot jk}/\pi_{\cdot j\cdot}
\]
does depend on $\tau^{\mathrm{ CG}}_{ij}$.

Now \textit{purge} $\pi_{ijk}$
of the cumbersome interaction by defining purged proportions
proportional to
\[
\pi_{ijk}^{**} = \pi_{ijk}/\tau^{\mathrm{ CG}}_{ij}\quad
\bigl(\mbox{i.e., }\pi_{ijk}^{*} = \pi_{ijk}^{**}/
\pi_{\cdots}^{**} \bigr).
\]
Actually,
\[
\pi_{ijk}^{*} = \eta^* \tau^{\mathrm{ C}}_i
\tau^{\mathrm{ G}}_j \tau^{\mathrm{ D}}_k
\tau^{\mathrm{ CD}}_{ik} \tau^{\mathrm{ GD}}_{jk}
\tau^{\mathrm{ CGD}}_{ijk}, \quad\eta^* = \eta/\pi_{\cdots}^{**}
\]
that is, the $\pi_{ijk}^*$ specify a model with all the same
parameters as before except that $\tau^{\mathrm{ CG}}_{ij}$ has
been replaced by 1.

Rates calculated from these adjusted proportions are now purged of the
$C\times G$
interaction but all other parameters are as before. Clogg noted the
fact that this
procedure is not the same as direct standardisation and defined a
variant,
\textit{marginal CG-purging}, which is equivalent to direct standardisation.

Purging was combined with further developments of additive
decomposition methods by \citet{Xie-1989} and \citet{Liao-1989} and was still
mentioned in the textbook by \citet{Powers-Xie-2008}, Section~4.6, but
seems otherwise to have played a modest part in recent decades. A very
interesting recent application is by \citet{Yamaguchi-2011}, who used
purging in counterfactual modelling of the mediation of the salary gap
between Japanese males and females by factors such as differential
educational attainment, use of part-time jobs and occupational
segregation.

\subsection*{Multiple Regression in Epidemiology}

By the early 1960s epidemiologists, in particular, cardiovascular
epidemiologists, were wrestling with the problem of \textit{multiple
causes}. It was clear that methods based solely on
cross-classification would have limited usefulness.
As put by \citet{truett:cornfield:kannel:1967}:

\begin{quote}
Thus, if 10 variables are under consideration, and each variable
is to be studied at only three levels, \ldots there would be 59,049
cells in the multiple cross-classification.
\end{quote}

\citet{Cornfield-1962} suggested the use of Fisher's discriminant
analysis to
deal with such problems. Although initially he considered only two
variables, he set out the idea more generally. This model assumes that
the vector of risk factor values is distributed, in
(incident) cases of
a disease and in subjects who remain disease free,
as multivariate normal variates with different means but equal
variance--covariance matrices. Reversing the conditioning by Bayes
theorem shows that the probability of disease given risk factors is
then given by the multiple logistic function.
The idea was investigated in more detail and for more risk
factors by \citet{truett:cornfield:kannel:1967} using data from the
12-year follow-up of subjects in the Framingham study. A clear concern was
that the multivariate normal assumption was clearly wrong in the
situations they were considering, which involved a mixture of
continuous and discrete risk factors. Despite this they demonstrated
that there was good correspondence between observed and expected risks
when subjects were classified according to deciles of the discriminant
function.

Truett et al. discussed the interpretation of the regression coefficients,
at some length, but did not remark on the connection with
multiplicative models and odds ratios, although Cornfield had, 15
years previously, established the approximate equivalence between the
odds ratio and a ratio of rates (see Section~\ref{sec:cc}). They did
note that the model is \textit{not} additive:

\begin{quote}
The relation between logit of risk and risk is illustrated in
Figure~1 \ldots a constant increase in the logit of risk does not
imply a constant increase in risk,
\end{quote}

\noindent and preferred to present the coefficients of the multiple logistic
function as multiples of the standard deviation of the corresponding
variable. They did, however, make it clear that these coefficients
represented an estimate of the effect of each risk factor \textit{after
holding all others constant}. They singled out the effect of weight
in this discussion:

\begin{quote}
The relative unimportance of weight as a risk factor \ldots when
all other risk factors are simultaneously considered is
noteworthy. This is not inconsistent with the possibility that a
reduction in weight would, by virtues of its effect on other risk
factors, for example, cholesterol, have important effects on the risk
of CHD.
\end{quote}

Finally, they noted that the model assumes the effect of each risk
factor to be independent of the levels of other risk factors, and
noted that first order interactions could be studied by relaxing the
assumption of equality of the variance--covariance matrices.

The avoidance of the assumption of multivariate normality in the
logistic model was achieved by use of the method of maximum
likelihood. In the epidemiological literature, this
is usually credited to \citet{Walker-Duncan-1967} who used a likelihood
based on conditioning on the values, $x$, of the risk
factors, and computing maximum likelihood estimates using the same
iteratively reweighted least squares algorithm proposed by
\citet{dyke:patterson:52}.
However, use of maximum likelihood in such models
had also been anticipated by \citet{Cox-1958}, although he had advocated
conditioning both on the observed set of risk factors, $x$, and on the
observed values of the disease status indicators, $y$. This is the
method, now known as ``conditional'' logistic regression, which
is important in the analysis of closely matched case--control
studies. Like Truett et al., Walker and Duncan gave little attention to
interpretation of the regression coefficients, save for advocating
standardization to SD units in an attempt to demonstrate the relative
importance of different factors. The main focus seems to have been in
risk prediction given multiple risk factors. \citet{Cox-1958},
however, discussed the
interpretation of the regression coefficient of
a dichomotous variable as a log odds ratio, even applying this to an
example, cited by \citet{Cornfield-1956},
concerning smoking and lung cancer in a survey of physicians.

A limitation of logistic regression for the analysis of follow-up
studies is the necessity to consider, as did
\citet{truett:cornfield:kannel:1967}, a fixed period of follow-up. A further
rationalization of analytical methods in epidemiology followed from
the realization that such studies generate right-censored,
and left-truncated, \textit{survival data}. \citet{mantel:66} pioneered
the modern approach to such problems,
noting that such data can be treated as if each
subject undergoes a series of Bernoulli trials (of very short
duration). He suggested, therefore, that
the comparison of survival between two groups could
be treated as an analysis of a $2 \times2\times K$ table in which the
$K$ ``trials'' are defined by the time points at which deaths occurred
in the study (other time points being uninformative). In his famous
paper, \citet{Cox-1972}, described a regression generalization of this
idea, in which the instantaneous risk, or ``hazard'', is predicted by
a log-linear regression model so that effects of each risk factor may
be expressed as hazard ratios. Over subsequent decades this theory was
further extended to encompass many types of event history data. See
\citet{andeborg} for a comprehensive review.

\subsection*{Confounder Scores and Propensity Scores}
Miettinen (\citeyear{Miettinen-1976}) put forward an alternative
proposal for dealing with
multiple confounders. It was motivated by three shortcomings he
identified in the multivariate methods then available:
\begin{longlist}[1.]
\item[1.] they (discriminant analysis in particular)
relied on very dubious assumptions,
\item[2.] they (logistic regression) were computationally demanding by
the standards then applying, and
\item[3.] they were poorly understood by substantive scientists.
\end{longlist}
His proposal was to carry out a preliminary, perhaps crude,
multivariate analysis from which could be computed a ``confounder
score''. This score could then be treated as a single confounder and
dealt with by conventional stratification methods. He suggested two
ways of computing the confounder score. An \textit{outcome function} was
computed by an initial regression (or discriminant function) analysis
of the
disease outcome variable on all of the confounders plus the exposure
variable of interest, then calculating the score for a fixed value of
exposure so that it depended solely on confounders. Alternatively, an
\textit{exposure function} could be computed by interchanging the
roles of outcome and exposure variables, regressing exposure on
confounders plus outcome.

\citet{Rosenbaum-Rubin-1983} later put forward a superficially similar
proposal to the use of Miettinen's exposure function. By analogy with
randomized experiments, they defined a \textit{balancing score} as a
function of potential confounders such that exposure and confounders
are conditionally independent given the balancing score. Stratification
by such
a score would then simulate a randomized experiment within each
stratum. They further demonstrated, for a binary exposure,
that the coarsest possible balancing score is the \textit{propensity
score}, the probability of exposure conditional upon confounders,
which can be estimated by logistic regression. Note that, unlike
Miettinen's exposure score, the outcome variable is not included in
this regression. The impact of estimation of the nuisance
parameters of the propensity score upon the test of exposure effect
was later explored by \citet{Rosenbaum-1984}. \citet{Hansen-2008}
later showed that a balancing score is also provided by the
``prognostic analogue'' to the propensity score
which is to Miettinen's outcome
function as the propensity score is to his exposure function, that is, the
exposure variable is omitted when calculating the prognostic score.

Given this later work on balancing scores, it is interesting to note
that Miettinen discussed at some length why he believed it necessary
to include the ``conditioning variable'' (either the exposure of
interest or the outcome variable) when computing the coefficients of
the confounder score, noting that the need for this was ``puzzling to some
epidemiologists''. His argument comes down to the requirement to
obtain an (approximately) unbiased estimate of the conditional
odds ratio for exposure versus outcome; omission of the conditioning
variable means that the confounder score potentially contains a component
related to only one of the two variables of interest and, owing to
noncollapsibility of the odds ratio, this leads to a biased
estimate of the conditional effect. Unfortunately, as demonstrated by
\citet{Pike-etal.-1979}, Miettinen's proposal for
correcting this bias comes at the cost of
inflation of the type~1 error rate for the hypothesis test for an
exposure effect. To demonstrate this, consider a logistic regression
of an outcome, $y$, on an exposure of interest, $x$, and multiple
confounders, $z$. Miettinen proposed to first compute a confounder
score $s = \hat{\gamma}^{\scriptsize T} z$, where $\hat{\gamma}$ are
the coefficients of $z$ in the logistic regression of $x$ on $y$ and
$z$, and then to fit the logistic regression of $y$ on $x$ and
$s$. While this regression yields an identical coefficient for $x$ as
the full logistic regression of $y$ on $x$ and $z$, and has the same
maximized likelihood, in the test for exposure effect this likelihood
is compared with the likelihood for the regression of $y$ on $s$
which, in general, will be rather less than that for $y$ on $z$---the
correct comparison point. Thus, the likelihood ratio test in
Miettinen's procedure will be inflated.

Rosenbaum and Rubin circumvented the estimation
problem posed by omission of the conditioning variable when
calculating balancing scores by estimating a \textit{marginal}
causal effect using direct standardization with appropriate population
weights. Equivalently, inverse probability weights based on the
propensity score can be used.

Owing to the focus on conditional measures of effect,
the propensity score approach was little used in epidemiology
during the latter part of the 20{th}
century. However, the method has gained considerably in
popularity over the last decade. For a recent case study of treatment
effect estimation using propensity score and regression
methods, see \citet{Kurth-etal.-2006}. They emphasised that,
as in classical direct standardization, precise identification of the
target population is important when treatment effects are nonuniform.

\subsection*{Time-Dependent Confounding}

Cox's life table regression model provided an exceedingly general
approach to modelling the probability of a failure event conditional
upon exposure or treatment variables and upon extraneous covariates or
confounders, the mathematical development extending quite naturally to
allow for variation of such variables over time. However, shortly
after its publication, \citet{kalbpren}, pages 124--126, pointed out a
serious difficulty in dealing with ``internal'' (endogenous)
time-dependent covariates.
Referring to the role of variables such as the general
condition of patients in therapeutic trials which may lie on the
causal path between earlier treatment and later outcome and,
therefore,
carry part
of the causal treatment effect, they wrote:

\begin{quote}
A censoring scheme that depends on the level of a time dependent
covariate $z(t)$ (e.g., general condition) is \ldots not independent
if $z(t)$ is not included in the model. One way to circumvent this
is to include $z(t)$ in the model, but this may mask treatment
differences of interest.
\end{quote}

Put another way, to ignore such a variable in the analysis
is to disregard its confounding
effect, but its inclusion
in the conditional probability model could
obscure some of the true causal effect of treatment.

While Kalbfleisch and Prentice had identified a fundamental problem
with the conditional approach to confounder adjustment, they offered no
convincing remedy. This was left to \citet{Robins-1986}. In this and
later papers Robins addressed the ``healthy worker'' effect in
epidemiology--essentially the same problem identified by
Kalbfleisch and Prentice.
Robins proposed two lines of attack which we may classify as
``marginal'' and ``conditional'' in keeping with a distinction that has
come up throughout our exposition. The original approach was
``g-computation'', which may be loosely conceived as sequential
prediction of ``what would have happened'' under various specified
externally imposed ``treatments'' and thus generalizes (direct)
standardisation, basically a marginal approach. A~further development
in this direction was inverse probability weighting of marginal
structural models, that is, models for the counterfactual outcomes
(\citep{Robins-etal.-2000}). Here, the essential idea is to estimate a marginal
treatment effect in a population in which the association between
treatment (exposure) and the time-dependent covariate is removed.
\citet{Sato-Matsuyama-2003} and \citet{Vansteelandt-Keiding-2011} gave brief
discussions of the relationship between g-computation, inverse
probability weighting and classical standardisation in the simplest
(nonlongitudinal) situation. On the other hand, the approach via
structural nested models [e.g., \citet{Robins-Tsiatis-1992},
\citet{Robins-etal.-1992}]
focusses on the effect of a ``blip'' of exposure at time $t$ conditional on
treatments and covariate values before $t$. In the latter models,
time-varying effect modification may be studied. See the recent
tutorial surveys by \citet{Robins-Hernan-2009} or \citet{Daniel-etal.-2013}
for details.

\section{Prediction and Transportability}
\label{sec:transport}
We saw that in the National Halothane Study standardisation methods
were used analytically, in order to control for confounders strictly
within the frame of the concrete study. The general verdict in the
emerging computer age regarding this use of standardisation was
negative, as
formulated by \citet{Fienberg-1975}, in a discussion of a careful and
detailed survey on observational studies by \citet{McKinlay-1975}:

\begin{quote}
The reader should be aware that standardization is basically a
descriptive technique that has been made obsolete, for most of the
purposes to which it has traditionally been put, by the ready
availability of computer programs for loglinear model analysis of
multidimensional contingency tables.
\end{quote}

However, the original use of standardization not only had this
analytical ambition, it also aimed at obtaining meaningful
generalizations to other populations---or other circumstances in the
original population. Before we sketch the recovery since the early
1980s of this aspect of standardization, it is useful to record the
attitude to generalization by influential epidemiologists back then.
\citeauthor{Miettinen-1985} [(\citeyear{Miettinen-1985}), page 47] in his long-awaited text-book, wrote:

\begin{quote}
In science the generalization from the actual study experience is not
made to a population of which the study experience is a sample in a
technical sense of probability sampling \ldots In science the
generalization is from the actual study experience to the abstract,
with no referent in place or time,
\end{quote}

\noindent and thus did not focus on specific recommendations as to how to
predict precisely what might happen under different concrete
circumstances. A similar attitude was voiced by \citeauthor{Rothman-1986}
[(\citeyear{Rothman-1986}),
page 95]
in the first edition of \textit{Modern Epidemiology}, and essentially repeated
in the following editions of this central reference work
[\citet{Rothman-Greenland-1998}, pages 133--134,
\citet{Rothman-etal.-2008}, pages 146--147]. The
immediate consequence of this attitude would be that all that we need
are the parameters in the fitted statistical model and assurance that
no bias is present in the genesis of the concretely analyzed data.

However, as we have seen, \citet{Clogg-1978} (and later) had
felt a need for interpreting the log-linear models in terms of their
consequences for summary tables. \citet{Freeman-Holford-1980} wrote a
useful guide to the new situation for survey analysis where the
collected data had been analyzed using the new statistical
models. They concluded that much favoured keeping the reporting to the
model parameters: these would then be available to other analysts for
comparative purposes, the model fit was necessary to check for
interactions (including possibly identifying a model where there is no
interaction). But,

\begin{quote}
in many settings these advantages are overshadowed by the dual
requirements for simplicity of presentation and immediacy of
interpretation,
\end{quote}

\noindent and \citet{Freeman-Holford-1980} therefore gave specific
instructions on how to
calculate ``summary rates'' for the total population or other
populations. The main requirement for validity of such calculations is
that there is no interaction between population and composition.

Interestingly, an influential contribution in 1982 came from a
rather different research environment: the well-established
agricultural statisticians P.~W. Lane and J.~A. Nelder
(\citep{Lane-Nelder-1982}). In a special issue of \textit{Biometrics}
on the theme
``the analysis of covariance'', they wrote a short note
containing several
germs of the later so important potential outcome view underlying
modern causal
inference, and placed the good old (direct) standardisation technique
right in
the middle of it.

Their view was that the purpose of a statistical analysis such as
analysis of covariance
is not only to estimate parameters, but also to make what they called
\textit{predictions}:

\begin{quote}
An essential feature is the division into effects of interest and
effects for which
adjustment is required. \ldots
For example, a typical prediction from a variety trial is the yield
that would
have been obtained from a particular variety if it had been grown over
the whole
experimental area. When a covariate exists the adjusted treatment mean
can be thought
of as the prediction of the yield of that variety grown over the whole
experimental
area with the covariate fixed at its mean value.
\ldots
The predictions here are not of future events but rather of what would
have happened
in the experiment if other conditions had prevailed. In fact no
variety would have
been grown over the whole experimental area nor would the covariate
have been constant.
\end{quote}

Lane and Nelder proposed to use the term \textit{predictive margin}
for such means,
avoiding the term ``population treatment mean'' suggested by
\citet{Searle-etal.-1980}
to replace the cryptic SAS-output term ``least square means''.
Lane and Nelder
emphasised that these means might either be

\begin{quote}
\textit{conditional} on the value we take as standard for the covariate
\end{quote}

\noindent or

\begin{quote}
\textit{marginal} to the observed distribution of covariate values,
\end{quote}

\noindent and Lane and Nelder went on to explain to this new audience (including
agricultural
statisticians) that there exist many other possibilities for choice of
standard.

We find it interesting that Lane and Nelder used the occasion of the
special issue of \textit{Biometrics}
on analysis of covariance to point out the \textit{similarities} to
standardisation,
and to phrase their ``prediction'' in much similar terms as the later causal
analysis would do. Of course, it should be remembered that Lane and
Nelder manoeuvered
within the comfortable framework of randomised field trials.
\citeauthor{Rothman-etal.-2008} [(\citeyear{Rothman-etal.-2008}), page 386 ff.]
described how these ideas have developed into what is now
termed \textit{regression standardisation.}

\subsection*{An Example: Cancer Trends}

A severe practical limitation of the modelling approach is that the
model must encompass all the data to be compared. However, many
official statistics are published explicitly to allow comparisons with
other published series. Even within a single publication it may be
inappropriate to fit a single large and complex model across the
entire data set.

An example of the latter situation is the I.A.R.C. monograph on Trends
in Cancer Incidence and Mortality (\citep{Coleman-etal.-1993}). The
primary aim of this monograph was to estimate cancer trends across the
population-based cancer registries throughout the world and this was
addressed by fitting age-period-cohort models to the data from each
registry. But comparisons of rates between registries at specific time
points were also required and, since the age structures of different
registries differed markedly, direct standardisation to the world
population, ages 30--74, was used. However, some of the cancers
considered were rare and this exposes a problem with the method of
direct standardization---that it can be very inefficient when the
standard population differs markedly from that of the test group. The
authors therefore chose to apply direct standardisation to the \emph
{fitted} rates from the age-period-cohort models.

\subsection*{Transportability Across Studies}

Pearl and
Barenboim (\citeyear{Pearl-Barenboim-2012}) noted that precise
conditions for
applying concrete results obtained in a study environment to another
``target'' environment,

\begin{quote}
remarkably\ldots have not received systematic formal treatment\ldots
The standard literature on this topic \ldots consists primarily of
``threats'', namely verbal narratives of what can go wrong when we try
to transport results from one study to another\ldots
Rarely do we find an analysis of ``licensing assumptions'', namely,
formal and transparent conditions under which the transport of results
across differing environments or populations is licensed from first
principles.
\end{quote}

After further outlining the strong odds against anyone who dares
formulate such conditions, Pearl and Barenboim then set out to
propose one such formalism, based on the causal diagrams developed by
Pearl and colleagues over the last decades; cf. \citet{Pearl-2009}.

In the terminology of
Pearl and Barenboim, the method of direct standardisation, together
with the ``predictions'' of Lane and Nelder, is a \emph{transport
formula} and, as they state,

\begin{quote}
the choice of the proper transport formula depends on the causal
context in which population differences are embedded.
\end{quote}

Although a formal treatment of these issues is overdue, it has
been recognized in epidemiology for many years that
the concept of confounding cannot be defined solely in terms of a
third variable being related to both outcome and exposure of interest.
A landmark paper was
that of \citet{Simpson-1951} which dealt with the problem of
interpreting associations in three-way contingency tables. As we
saw in Section~\ref{sec:association}, although
``Simpson's paradox'' is widely regarded as synonymous with
Yule's paradox, Simpson's primary
concern was the role of the causal context in deciding whether
the conditional or marginal association between two of the three factors
in a table is
of primary interest. The point has been understood by many (if
not all) epidemiologists writing in the second half of the
20{th} century as, for example, is demonstrated by
the remark of \citet{truett:cornfield:kannel:1967}, cited in
Section~\ref{sec:models},
concerning interpretation of the coefficient of body weight in their
regression equation for coronary disease incidence.
However, as far as we can tell, the issue
does not seem to have concerned 19{th} century
writers; for example, no consideration seems to have been given to the
possibility that age differences between populations
could, in part, be a consequence of differences in ``the force of
mortality'' and, if so, the implication for age standardization.

\section{Conclusion}\label{sec:conclusion}

In the fields of scientific enquiry with which we are concerned here,
the causal effect of a treatment, or exposure, cannot be observed at
the individual level. Instead, the effect measures we use contrast the
distributions of responses in populations with differing
exposures, but in which the distributions of other factors do not
differ. In randomized studies, this equality of distribution of
extraneous factors is guaranteed by randomisation and causal effects
are simply measured. In observational studies, however, differences between
the distributions of relevant extraneous factors between exposure
groups (what epidemiologists call ``confounding'') is ubiquitous and we
must rely on the assumption of ``no unmeasured confounders'' to allow us
to estimate the causal effect.

In much recent work, the problem is
approached by postulating that each individual has a number of
potential responses, one for each possible exposure; only one of these
is observed, the other counterfactual responses being assumed to be
``missing at random'' given measured confounders. Alternatively, we can
restrict ourselves to dealing with observed outcomes, assuming that
the mechanism by which exposure was allocated in the
experiment of nature we have observed did not depend on unmeasured
confounders. The choice between these positions is philosophical
and, to the applied statistician, largely a matter of convenience. The
more serious concern, with which we have been largely concerned in
this review, is the choice of effect measure; we can choose to
contrast the marginal distribution of responses under equality of
distribution of extraneous factors or to contrast response
distributions which
condition on the values of these factors.

Standardisation grew up in response to obvious problems of
age-confounding in actuarial (18{th} century) and
demographic (19{th}
century) comparative studies of mortality. The simple intuitive
calculations
considered scenarios in which either the age distributions did not
differ (``direct'' standardisation) or
age-specific rates did not differ (``indirect'' standardisation) between
study groups. However, formal consideration of such indices as effect
measures came later, the contribution of \citet{Yule-1934} being
noteworthy.

There would probably be widespread agreement that describing causal effects
in relation to \textit{all} potential causes, as in the conditional
approach, must represent the most complete analysis of a data set and
we have
described how this approach developed throughout
the 20{th} century, starting
with the influential paper of \citet{Yule-1900}. This impressive paper
clearly described the proliferation of ``partial'' association measures
introduced by the conditional approach, and drew attention to the
consequent need to use measures which remain relatively
stable across subgroups. In later work \citet{Yule-1934} revisited
classical standardisation in terms of an average of conditional
(i.e., stratum-specific)
measures. Such early work sowed the seeds of the ``statistical''
approach based on formal probability models, leading eventually to the
widespread use of logistic
regression and proportional hazard (multiplicative intensity) models, the
contributions of \citet{cochran:54} and
\citet{mantel:haenszel:1959} providing important staging posts along the
way (even though the latter authors explicitly denied any reliance on a
model). By the end of the century such approaches dominated
epidemiology and biostatistics.

Toward the end of the 20{th} century, the use of marginal
measures of causal effects emerged from the counterfactual approach to
causal analysis
in the social sciences, the idea of propensity scores
(\citep{Rosenbaum-Rubin-1983}) being particularly influential. However,
these methods only found their way into mainstream biostatistics
when applications arose for which conventional conditional probability
models were not well suited, the foremost of which being the problem of
time-dependent confounding. Whereas, for simple problems, the
parameters of logistic and multiplicative intensity models have an
interpretation as measures of (conditional) causal effects,
\citet{kalbpren} noted that this ceased to be the case in the presence
of time-dependent confounding. While the statistical modelling
approach can be extended by joint modelling of the event history and
confounder trajectories, such models are complex and, again, causal
effects do not correspond with model parameters and would need to be
estimated by simulations based on the fitted model. Such models
continue to be
studied [see, e.g., \citet{Henderson-Diggle-Dobson-2000}], but,
although it could be argued that these offer the opportunity for a more
detailed understanding of the nature of causal effects in this setting,
the simpler marginal approaches pioneered by
\citet{Robins-1986} are more attractive in most applications.
The success of this latter
approach in offering a solution to a previously intractable problem
encouraged biostatisticians to further explore methods for estimation
of marginal causal effects; for example, propensity scores are now
widely used in this literature.

So where are we today? Both approaches have strengths and
weaknesses. The conditional modelling approach relies on the
assumption of homogeneity of effect across subgroups or, when this
fails to hold, to a multiplicity of effect measures. The
marginal approach, while seemingly less reliant on such assumptions,
encounters the same issue when considering the transportability of
effect measures to different populations.

\section*{Acknowledgments}
This work grew out of an invited paper by David Clayton in the session
``Meaningful Parametrizations'' at the International Biometric
Conference in Freiburg in 2002. The authors are grateful to the
reviewers for comments and detailed
suggestions on earlier drafts.
Important parts of Niels Keiding's
work were done on study leaves at the MRC Biostatistics Unit,
Cambridge, October--November 2009 and as Visiting Senior Research
Fellow at Jesus College, Oxford, Hilary Term 2012; he is most grateful
for the hospitality at these institutions.
David Clayton was supported
by a Wellcome Trust Principal Research Fellowship until his retirement
in March, 2012.


%


\begin{thebibliography}{146}

\bibitem[\protect\citeauthoryear{Aalen}{1978}]{aalen:78}
%
\begin{barticle}[mr]
\bauthor{\bsnm{Aalen},~\bfnm{Odd}\binits{O.}}
(\byear{1978}).
\btitle{Nonparametric inference for a family of counting processes}.
\bjournal{Ann. Statist.}
\bvolume{6}
\bpages{701--726}.
\bid{issn={0090-5364}, mr={0491547}}
\bptok{imsref}%
\end{barticle}
%
\endbibitem

\bibitem[\protect\citeauthoryear{Aldrich}{1995}]{Aldrich-1995}
%
\begin{barticle}[auto:STB|2013/12/09|07:59:19]
\bauthor{\bsnm{Aldrich},~\bfnm{J.}\binits{J.}}
(\byear{1995}).
\btitle{Correlations genuine and spurious in Pearson and Yule}.
\bjournal{Statist. Sci.}
\bvolume{10}
\bpages{364--376}.
\bptok{imsref}%
\end{barticle}
%
\endbibitem

\bibitem[\protect\citeauthoryear{Andersen et~al.}{1993}]{andeborg}
%
\begin{bbook}[mr]
\bauthor{\bsnm{Andersen},~\bfnm{Per~Kragh}\binits{P.~K.}},
\bauthor{\bsnm{Borgan},~\bfnm{{\O}rnulf}\binits{{\O}.}},
\bauthor{\bsnm{Gill},~\bfnm{Richard~D.}\binits{R.~D.}} \AND
\bauthor{\bsnm{Keiding},~\bfnm{Niels}\binits{N.}}
(\byear{1993}).
\btitle{Statistical Models Based on Counting Processes}.
\bpublisher{Springer}, \blocation{New York}.
\bid{doi={10.1007/978-1-4612-4348-9}, mr={1198884}}
\bptok{imsref}%
\end{bbook}
%
\endbibitem

\bibitem[\protect\citeauthoryear{Bartlett}{1935}]{Bartlett-1935}
%
\begin{barticle}[auto:STB|2013/12/09|07:59:19]
\bauthor{\bsnm{Bartlett},~\bfnm{M.~S.}\binits{M.~S.}}
(\byear{1935}).
\btitle{Contingency table interactions}.
\bjournal{Suppl. J. Roy. Stat. Soc.}
\bvolume{2}
\bpages{248--252}.
\bptok{imsref}%
\end{barticle}
%
\endbibitem

\bibitem[\protect\citeauthoryear{Bellhouse}{2008}]{Bellhouse-2008}
%
\begin{barticle}[auto:STB|2013/12/09|07:59:19]
\bauthor{\bsnm{Bellhouse},~\bfnm{D.}\binits{D.}}
(\byear{2008}).
\btitle{Review of ``{Disciplining Statistics: Demography and Vital Statistics
in France and England, 1830--1885}'' by {L. S}chweber}.
\bjournal{Hist. Math.}
\bvolume{35}
\bpages{249--252}.
\bptok{imsref}%
\end{barticle}
%
\endbibitem

\bibitem[\protect\citeauthoryear{Belson}{1956}]{Belson-1956}
%
\begin{barticle}[auto:STB|2013/12/09|07:59:19]
\bauthor{\bsnm{Belson},~\bfnm{W.~A.}\binits{W.~A.}}
(\byear{1956}).
\btitle{A technique for studying the effects of a television broadcast}.
\bjournal{Appl. Stat.}
\bvolume{5}
\bpages{195--202}.
\bptok{imsref}%
\end{barticle}
%
\endbibitem

\bibitem[\protect\citeauthoryear{Berkson}{1946}]{Berkson-1946}
%
\begin{barticle}[auto:STB|2013/12/09|07:59:19]
\bauthor{\bsnm{Berkson},~\bfnm{J.}\binits{J.}}
(\byear{1946}).
\btitle{Limitations of the application of fourfold table analysis to hospital
data}.
\bjournal{Biometrics Bull.}
\bvolume{2}
\bpages{47--53}.
\bptok{imsref}%
\end{barticle}
%
\endbibitem

\bibitem[\protect\citeauthoryear{Birch}{1963}]{Birch-1963}
%
\begin{barticle}[mr]
\bauthor{\bsnm{Birch},~\bfnm{M.~W.}\binits{M.~W.}}
(\byear{1963}).
\btitle{Maximum likelihood in three-way contingency tables}.
\bjournal{J. R. Stat. Soc. Ser. B Stat. Methodol.}
\bvolume{25}
\bpages{220--233}.
\bid{issn={0035-9246}, mr={0168065}}
\bptok{imsref}%
\end{barticle}
%
\endbibitem

\bibitem[\protect\citeauthoryear{Birch}{1964}]{Birch-1964}
%
\begin{barticle}[mr]
\bauthor{\bsnm{Birch},~\bfnm{M.~W.}\binits{M.~W.}}
(\byear{1964}).
\btitle{The detection of partial association. {I}. {T}he {$2\times2$} case}.
\bjournal{J. R. Stat. Soc. Ser. B Stat. Methodol.}
\bvolume{26}
\bpages{313--324}.
\bid{issn={0035-9246}, mr={0176562}}
\bptok{imsref}%
\end{barticle}
%
\endbibitem

\bibitem[\protect\citeauthoryear{Bishop}{1969}]{Bishop-1969}
%
\begin{barticle}[auto:STB|2013/12/09|07:59:19]
\bauthor{\bsnm{Bishop},~\bfnm{Y.~M.~M.}\binits{Y.~M.~M.}}
(\byear{1969}).
\btitle{Full contingency tables, logits, and split contingency tables}.
\bjournal{Biometrics}
\bvolume{25}
\bpages{383--399}.
\bptok{imsref}%
\end{barticle}
%
\endbibitem

\bibitem[\protect\citeauthoryear{Bishop}{1971}]{Bishop-1971}
%
\begin{barticle}[auto:STB|2013/12/09|07:59:19]
\bauthor{\bsnm{Bishop},~\bfnm{Y.~M.~M.}\binits{Y.~M.~M.}}
(\byear{1971}).
\btitle{Effects of collapsing multidimensional contingency tables}.
\bjournal{Biometrics}
\bvolume{27}
\bpages{545--562}.
\bptok{imsref}%
\end{barticle}
%
\endbibitem

\bibitem[\protect\citeauthoryear{Bishop, Fienberg and
Holland}{1975}]{Bishop-Fienberg-Holland-1975}
%
\begin{bbook}[mr]
\bauthor{\bsnm{Bishop},~\bfnm{Yvonne M.~M.}\binits{Y.~M.~M.}},
\bauthor{\bsnm{Fienberg},~\bfnm{Stephen~E.}\binits{S.~E.}} \AND
\bauthor{\bsnm{Holland},~\bfnm{Paul~W.}\binits{P.~W.}}
(\byear{1975}).
\btitle{Discrete Multivariate Analysis: Theory and Practice}.
\bpublisher{MIT Press}, \blocation{Cambridge, MA}.
\bid{mr={0381130}}
\bptok{imsref}%
\end{bbook}
%
\endbibitem

\bibitem[\protect\citeauthoryear{Borgan}{1984}]{Borgan-1984}
%
\begin{barticle}[mr]
\bauthor{\bsnm{Borgan},~\bfnm{{\O}rnulf}\binits{{\O}.}}
(\byear{1984}).
\btitle{Maximum likelihood estimation in parametric counting process models,
with applications to censored failure time data}.
\bjournal{Scand. J. Stat.}
\bvolume{11}
\bpages{1--16}.
\bid{issn={0303-6898}, mr={0743234}}
\bptok{imsref}%
\end{barticle}
%
\endbibitem

\bibitem[\protect\citeauthoryear{Breslow and Day}{1975}]{Breslow-Day-1975}
%
\begin{barticle}[auto:STB|2013/12/09|07:59:19]
\bauthor{\bsnm{Breslow},~\bfnm{N.~E.}\binits{N.~E.}} \AND
\bauthor{\bsnm{Day},~\bfnm{N.~E.}\binits{N.~E.}}
(\byear{1975}).
\btitle{Indirect standardization and multiplicative models for rates, with
reference to age adjustment of cancer incidence and relative frequency data}.
\bjournal{J. Chron. Dis.}
\bvolume{28}
\bpages{289--303}.
\bptok{imsref}%
\end{barticle}
%
\endbibitem

\bibitem[\protect\citeauthoryear{Breslow et~al.}{1983}]{Breslow-etal.-1983}
%
\begin{barticle}[auto:STB|2013/12/09|07:59:19]
\bauthor{\bsnm{Breslow},~\bfnm{N.~E.}\binits{N.~E.}},
\bauthor{\bsnm{Lubin},~\bfnm{J.~H.}\binits{J.~H.}},
\bauthor{\bsnm{Marek},~\bfnm{P.}\binits{P.}} \AND
\bauthor{\bsnm{Langholz},~\bfnm{B.}\binits{B.}}
(\byear{1983}).
\btitle{Multiplicative models and cohort analysis}.
\bjournal{J. Amer. Statist. Assoc.}
\bvolume{78}
\bpages{1--12}.
\bptok{imsref}%
\end{barticle}
%
\endbibitem

\bibitem[\protect\citeauthoryear{Bunker et~al.}{1969}]{Bunker-etal.-1969}
%
\begin{bmisc}[auto:STB|2013/12/09|07:59:19]
\bauthor{\bsnm{Bunker},~\bfnm{J.}\binits{J.}},
\bauthor{\bsnm{Forrest},~\bfnm{W.}\binits{W.}},
\bauthor{\bsnm{Mosteller},~\bfnm{F.}\binits{F.}} \AND
\bauthor{\bsnm{Vandam},~\bfnm{L.}\binits{L.}}
(\byear{1969}).
\bhowpublished{\textit{The National Halothane Study}. National
Institute of
General Medical Sciences}.
\bptok{imsref}%
\end{bmisc}
%
\endbibitem

\bibitem[\protect\citeauthoryear{Chadwick}{1844}]{Chadwick-1844}
%
\begin{barticle}[auto:STB|2013/12/09|07:59:19]
\bauthor{\bsnm{Chadwick},~\bfnm{E.}\binits{E.}}
(\byear{1844}).
\btitle{On the best modes of representing accurately, by statistical returns,
the duration of life, and the pressure and progress of the causes of
mortality amongst different classes of the community, and amongst the
populations of different districts and countries}.
\bjournal{J. Stat. Soc. London}
\bvolume{7}
\bpages{1--40}.
\bptok{imsref}%
\end{barticle}
%
\endbibitem

\bibitem[\protect\citeauthoryear{Chevan and
Sutherland}{2009}]{Chevan-Sutherland-2009}
%
\begin{barticle}[auto:STB|2013/12/09|07:59:19]
\bauthor{\bsnm{Chevan},~\bfnm{A.}\binits{A.}} \AND
\bauthor{\bsnm{Sutherland},~\bfnm{M.}\binits{M.}}
(\byear{2009}).
\btitle{Revisiting Das Gupta: Refinement and extension of
standardization and
decomposition}.
\bjournal{Demography}
\bvolume{46}
\bpages{429--449}.
\bptok{imsref}%
\end{barticle}
%
\endbibitem

\bibitem[\protect\citeauthoryear{Clayton and Hills}{1993}]{clayton:hills:93}
%
\begin{bbook}[auto:STB|2013/12/09|07:59:19]
\bauthor{\bsnm{Clayton},~\bfnm{D.}\binits{D.}} \AND
\bauthor{\bsnm{Hills},~\bfnm{M.}\binits{M.}}
(\byear{1993}).
\btitle{Statistical Models in Epidemiology}.
\bpublisher{Oxford Univ. Press}, \blocation{Oxford}.
\bptok{imsref}%
\end{bbook}
%
\endbibitem

\bibitem[\protect\citeauthoryear{Clogg}{1978}]{Clogg-1978}
%
\begin{barticle}[pbm]
\bauthor{\bsnm{Clogg},~\bfnm{C.~C.}\binits{C.~C.}}
(\byear{1978}).
\btitle{Adjustment of rates using multiplicative models}.
\bjournal{Demography}
\bvolume{15}
\bpages{523--539}.
\bid{issn={0070-3370}, pmid={738478}}
\bptok{imsref}%
\end{barticle}
%
\endbibitem

\bibitem[\protect\citeauthoryear{Clogg and Eliason}{1988}]{Clogg-Eliason-1988}
%
\begin{barticle}[auto:STB|2013/12/09|07:59:19]
\bauthor{\bsnm{Clogg},~\bfnm{C.~C.}\binits{C.~C.}} \AND
\bauthor{\bsnm{Eliason},~\bfnm{S.~R.}\binits{S.~R.}}
(\byear{1988}).
\btitle{A flexible procedure for adjusting rates and proportions, including
statistical methods for group comparisons}.
\bjournal{Am. Sociol. Rev.}
\bvolume{53}
\bpages{267--283}.
\bptok{imsref}%
\end{barticle}
%
\endbibitem

\bibitem[\protect\citeauthoryear{Cochran}{1954}]{cochran:54}
%
\begin{barticle}[mr]
\bauthor{\bsnm{Cochran},~\bfnm{William~G.}\binits{W.~G.}}
(\byear{1954}).
\btitle{Some methods for strengthening the common {$\chi^2$} tests}.
\bjournal{Biometrics}
\bvolume{10}
\bpages{417--451}.
\bid{issn={0006-341X}, mr={0067428}}
\bptok{imsref}%
\end{barticle}
%
\endbibitem

\bibitem[\protect\citeauthoryear{Cochran}{1969}]{Cochran-1969}
%
\begin{barticle}[auto:STB|2013/12/09|07:59:19]
\bauthor{\bsnm{Cochran},~\bfnm{W.~G.}\binits{W.~G.}}
(\byear{1969}).
\btitle{Use of covariance in observational studies}.
\bjournal{Appl. Stat.}
\bvolume{18}
\bpages{270--275}.
\bptok{imsref}%
\end{barticle}
%
\endbibitem

\bibitem[\protect\citeauthoryear{Coleman et~al.}{1993}]{Coleman-etal.-1993}
%
\begin{bbook}[auto:STB|2013/12/09|07:59:19]
\bauthor{\bsnm{Coleman},~\bfnm{M.~P.}\binits{M.~P.}},
\bauthor{\bsnm{Est{\`e}ve},~\bfnm{J.}\binits{J.}},
\bauthor{\bsnm{Damiecki},~\bfnm{P.}\binits{P.}},
\bauthor{\bsnm{Arslan},~\bfnm{A.}\binits{A.}} \AND
\bauthor{\bsnm{Renard},~\bfnm{H.}\binits{H.}}
(\byear{1993}).
\btitle{Trends in Cancer Incidence and Mortality. No. 121 in IARC Scientific Publications}.
\bpublisher{International Agency for
Research on Cancer},
\blocation{Lyon}.
\bptok{imsref}%
\end{bbook}
%
\endbibitem

\bibitem[\protect\citeauthoryear{Cornfield}{1951}]{cornfield:1951}
%
\begin{barticle}[pbm]
\bauthor{\bsnm{Cornfield},~\bfnm{J.}\binits{J.}}
(\byear{1951}).
\btitle{A method of estimating comparative rates from clinical data;
applications to cancer of the lung, breast, and cervix}.
\bjournal{J. Natl. Cancer Inst.}
\bvolume{11}
\bpages{1269--1275}.
\bid{issn={0027-8874}, pmid={14861651}}
\bptok{imsref}%
\end{barticle}
%
\endbibitem

\bibitem[\protect\citeauthoryear{Cornfield}{1956}]{Cornfield-1956}
%
\begin{binproceedings}[mr]
\bauthor{\bsnm{Cornfield},~\bfnm{Jerome}\binits{J.}}
(\byear{1956}).
\btitle{A statistical problem arising from retrospective studies}.
In \bbooktitle{Proceedings of the {T}hird {B}erkeley {S}ymposium on
{M}athematical {S}tatistics and {P}robability, 1954--1955, Vol. {IV}}
(\beditor{\bfnm{J.}\binits{J.}~\bsnm{Neyman}}, ed.)
\bpages{135--148}.
\bpublisher{Univ. California Press}, \blocation{Berkeley, CA}.
\bid{mr={0084935}}
\bptok{imsref}%
\end{binproceedings}
%
\endbibitem

\bibitem[\protect\citeauthoryear{Cornfield}{1962}]{Cornfield-1962}
%
\begin{barticle}[pbm]
\bauthor{\bsnm{Cornfield},~\bfnm{J.}\binits{J.}}
(\byear{1962}).
\btitle{Joint dependence of risk of coronary heart disease on serum cholesterol
and systolic blood pressure: A discriminant function analysis}.
\bjournal{Fed. Proc.}
\bvolume{21}
\bpages{58--61}.
\bid{issn={0014-9446}, pmid={13881407}}
\bptok{imsref}%
\end{barticle}
%
\endbibitem

\bibitem[\protect\citeauthoryear{Cox}{1958}]{Cox-1958}
%
\begin{barticle}[mr]
\bauthor{\bsnm{Cox},~\bfnm{D.~R.}\binits{D.~R.}}
(\byear{1958}).
\btitle{The regression analysis of binary sequences}.
\bjournal{J. R. Stat. Soc. Ser. B Stat. Methodol.}
\bvolume{20}
\bpages{215--242}.
\bid{issn={0035-9246}, mr={0099097}}
\bptok{imsref}%
\end{barticle}
%
\endbibitem

\bibitem[\protect\citeauthoryear{Cox}{1972}]{Cox-1972}
%
\begin{barticle}[mr]
\bauthor{\bsnm{Cox},~\bfnm{D.~R.}\binits{D.~R.}}
(\byear{1972}).
\btitle{Regression models and life-tables}.
\bjournal{J. R. Stat. Soc. Ser. B Stat. Methodol.}
\bvolume{34}
\bpages{187--220}.
\bid{issn={0035-9246}, mr={0341758}}
\bptnote{check related}%
\bptok{imsref}%
\end{barticle}
%
\endbibitem

\bibitem[\protect\citeauthoryear{Daniel et~al.}{2013}]{Daniel-etal.-2013}
%
\begin{barticle}[mr]
\bauthor{\bsnm{Daniel},~\bfnm{R.~M.}\binits{R.~M.}},
\bauthor{\bsnm{Cousens},~\bfnm{S.~N.}\binits{S.~N.}},
\bauthor{\bsnm{De~Stavola},~\bfnm{B.~L.}\binits{B.~L.}},
\bauthor{\bsnm{Kenward},~\bfnm{M.~G.}\binits{M.~G.}} \AND
\bauthor{\bsnm{Sterne},~\bfnm{J.~A.~C.}\binits{J.~A.~C.}}
(\byear{2013}).
\btitle{Methods for dealing with time-dependent confounding}.
\bjournal{Stat. Med.}
\bvolume{32}
\bpages{1584--1618}.
\bid{doi={10.1002/sim.5686}, issn={0277-6715}, mr={3060620}}
\bptok{imsref}%
\end{barticle}
%
\endbibitem

\bibitem[\protect\citeauthoryear{Davidson and
Reid}{1926--1927}]{Davidson-Reid-1926-1927}
%
\begin{barticle}[auto:STB|2013/12/09|07:59:19]
\bauthor{\bsnm{Davidson},~\bfnm{A.}\binits{A.}} \AND
\bauthor{\bsnm{Reid},~\bfnm{A.}\binits{A.}}
(\byear{1926--1927}).
\btitle{On the calculation of rates of mortality}.
\bjournal{T. Fac. Actuaries}
\bvolume{11}
\bpages{183--232}.
\bptok{imsref}%
\end{barticle}
%
\endbibitem

\bibitem[\protect\citeauthoryear{Day}{1976}]{Day-1976}
%
\begin{bincollection}[auto:STB|2013/12/09|07:59:19]
\bauthor{\bsnm{Day},~\bfnm{N.~E.}\binits{N.~E.}}
(\byear{1976}).
\btitle{A new measure of age standardized incidence, the cumulative rate}.
In \bbooktitle{Cancer Incidence in Five Continents 3}
(\beditor{\bfnm{J.}\binits{J.}~\bsnm{Waterhouse}},
\beditor{\bfnm{C.}\binits{C.}~\bsnm{Muir}},
\beditor{\bfnm{P.}\binits{P.}~\bsnm{Correa}} \AND
\beditor{\bfnm{J.}\binits{J.}~\bsnm{Powell}}, eds.)
\bpages{443--445}.
\bpublisher{International Agency for Research on Cancer}, \blocation{Lyon}.
\bptok{imsref}%
\end{bincollection}
%
\endbibitem

\bibitem[\protect\citeauthoryear{Derrick}{1927}]{Derrick-1927}
%
\begin{barticle}[auto:STB|2013/12/09|07:59:19]
\bauthor{\bsnm{Derrick},~\bfnm{V.~P.~A.}\binits{V.~P.~A.}}
(\byear{1927}).
\btitle{Observations on (1) errors of age in the population statistics of
England and Wales, and (2) the changes in mortality indicated by the national
records}.
\bjournal{J. Inst. Actuaries}
\bvolume{58}
\bpages{117--159}.
\bptok{imsref}%
\end{barticle}
%
\endbibitem

\bibitem[\protect\citeauthoryear{Dyke and Patterson}{1952}]{dyke:patterson:52}
%
\begin{barticle}[auto:STB|2013/12/09|07:59:19]
\bauthor{\bsnm{Dyke},~\bfnm{G.}\binits{G.}} \AND
\bauthor{\bsnm{Patterson},~\bfnm{H.}\binits{H.}}
(\byear{1952}).
\btitle{Analysis of factorial arrangements when the data are proportions}.
\bjournal{Biometrics}
\bvolume{8}
\bpages{1--12}.
\bptok{imsref}%
\end{barticle}
%
\endbibitem

\bibitem[\protect\citeauthoryear{Edgeworth}{1917}]{Edgeworth-1917}
%
\begin{barticle}[auto:STB|2013/12/09|07:59:19]
\bauthor{\bsnm{Edgeworth},~\bfnm{F.~Y.}\binits{F.~Y.}}
(\byear{1917}).
\btitle{Review of ``Scope and Method of Statistics'' by Harald Westergaard}.
\bjournal{J. Roy. Stat. Soc.}
\bvolume{80}
\bpages{546--551}.
\bptok{imsref}%
\end{barticle}
%
\endbibitem

\bibitem[\protect\citeauthoryear{Elandt-Johnson}{1975}]{elandt-johnson:75}
%
\begin{barticle}[pbm]
\bauthor{\bsnm{Elandt-Johnson},~\bfnm{R.~C.}\binits{R.~C.}}
(\byear{1975}).
\btitle{Definition of rates: Some remarks on their use and misuse}.
\bjournal{Am. J. Epidemiol.}
\bvolume{102}
\bpages{267--271}.
\bid{issn={0002-9262}, pmid={1180251}}
\bptok{imsref}%
\end{barticle}
%
\endbibitem

\bibitem[\protect\citeauthoryear{Farr}{1859}]{Farr-1859}
%
\begin{bbook}[auto:STB|2013/12/09|07:59:19]
\bauthor{\bsnm{Farr},~\bfnm{W.}\binits{W.}}
(\byear{1859}).
\btitle{Letter to the Registrar General}.
\bpublisher{General Registrar Office}, \blocation{London}.
\bptok{imsref}%
\end{bbook}
%
\endbibitem

\bibitem[\protect\citeauthoryear{Fibiger}{1898}]{Fibiger-1898}
%
\begin{barticle}[auto:STB|2013/12/09|07:59:19]
\bauthor{\bsnm{Fibiger},~\bfnm{J.}\binits{J.}}
(\byear{1898}).
\btitle{Om serumbehandling af difteri}.
\bjournal{Hospitalstidende}
\bvolume{6}
\bpages{337--350}.
\bptok{imsref}%
\end{barticle}
%
\endbibitem

\bibitem[\protect\citeauthoryear{Fienberg}{1975}]{Fienberg-1975}
%
\begin{barticle}[auto:STB|2013/12/09|07:59:19]
\bauthor{\bsnm{Fienberg},~\bfnm{S.~E.}\binits{S.~E.}}
(\byear{1975}).
\btitle{Design and analysis of observational study---Comment}.
\bjournal{J. Amer. Statist. Assoc.}
\bvolume{70}
\bpages{521--523}.
\bptok{imsref}%
\end{barticle}
%
\endbibitem

\bibitem[\protect\citeauthoryear{Fisher}{1922}]{Fisher-1922}
%
\begin{barticle}[auto:STB|2013/12/09|07:59:19]
\bauthor{\bsnm{Fisher},~\bfnm{R.~A.}\binits{R.~A.}}
(\byear{1922}).
\btitle{On the interpretation of $\chi^2$ from contingency tables, and the
calculation of ${P}$}.
\bjournal{J. Roy. Stat. Soc.}
\bvolume{85}
\bpages{87--94}.
\bptok{imsref}%
\end{barticle}
%
\endbibitem

\bibitem[\protect\citeauthoryear{Freeman and
Holford}{1980}]{Freeman-Holford-1980}
%
\begin{barticle}[pbm]
\bauthor{\bsnm{Freeman},~\bfnm{D.~H.}\binits{D.~H.}} \AND
\bauthor{\bsnm{Holford},~\bfnm{T.~R.}\binits{T.~R.}}
(\byear{1980}).
\btitle{Summary rates}.
\bjournal{Biometrics}
\bvolume{36}
\bpages{195--205}.
\bid{issn={0006-341X}, pmid={7407310}}
\bptok{imsref}%
\end{barticle}
%
\endbibitem

\bibitem[\protect\citeauthoryear{Graubard, Rao and
Gastwirth}{2005}]{Graubard-etal.-2005}
%
\begin{barticle}[mr]
\bauthor{\bsnm{Graubard},~\bfnm{B.~I.}\binits{B.~I.}},
\bauthor{\bsnm{Rao},~\bfnm{R.~Sowmya}\binits{R.~S.}} \AND
\bauthor{\bsnm{Gastwirth},~\bfnm{Joseph~L.}\binits{J.~L.}}
(\byear{2005}).
\btitle{Using the {P}eters--{B}elson method to measure health care disparities
from complex survey data}.
\bjournal{Stat. Med.}
\bvolume{24}
\bpages{2659--2668}.
\bid{doi={10.1002/sim.2135}, issn={0277-6715}, mr={2196206}}
\bptok{imsref}%
\end{barticle}
%
\endbibitem

\bibitem[\protect\citeauthoryear{Greenland}{1982}]{Greenland-1982}
%
\begin{barticle}[pbm]
\bauthor{\bsnm{Greenland},~\bfnm{S.}\binits{S.}}
(\byear{1982}).
\btitle{Interpretation and estimation of summary ratios under heterogeneity}.
\bjournal{Stat. Med.}
\bvolume{1}
\bpages{217--227}.
\bid{issn={0277-6715}, pmid={7187095}}
\bptok{imsref}%
\end{barticle}
%
\endbibitem

\bibitem[\protect\citeauthoryear{Haberman}{1974}]{Haberman-1974}
%
\begin{bbook}[mr]
\bauthor{\bsnm{Haberman},~\bfnm{Shelby~J.}\binits{S.~J.}}
(\byear{1974}).
\btitle{The Analysis of Frequency Data: Statistical Research
Monographs, Vol.
IV}.
\bpublisher{The Univ. Chicago Press}, \blocation{Chicago, IL}.
\bid{mr={0408098}}
\bptok{imsref}%
\end{bbook}
%
\endbibitem

\bibitem[\protect\citeauthoryear{Haenszel, Shimkin and
Mantel}{1958}]{haenszel:shimkin:mantel:1958}
%
\begin{barticle}[pbm]
\bauthor{\bsnm{Haenszel},~\bfnm{W.}\binits{W.}},
\bauthor{\bsnm{Shimkin},~\bfnm{M.~B.}\binits{M.~B.}} \AND
\bauthor{\bsnm{Mantel},~\bfnm{N.}\binits{N.}}
(\byear{1958}).
\btitle{A~retrospective study of lung cancer in women}.
\bjournal{J. Natl. Cancer Inst.}
\bvolume{21}
\bpages{825--842}.
\bid{issn={0027-8874}, pmid={13599015}}
\bptok{imsref}%
\end{barticle}
%
\endbibitem

\bibitem[\protect\citeauthoryear{Hansen}{2008}]{Hansen-2008}
%
\begin{barticle}[mr]
\bauthor{\bsnm{Hansen},~\bfnm{Ben~B.}\binits{B.~B.}}
(\byear{2008}).
\btitle{The prognostic analogue of the propensity score}.
\bjournal{Biometrika}
\bvolume{95}
\bpages{481--488}.
\bid{doi={10.1093/biomet/asn004}, issn={0006-3444}, mr={2521594}}
\bptok{imsref}%
\end{barticle}
%
\endbibitem

\bibitem[\protect\citeauthoryear{Henderson, Diggle and
Dobson}{2000}]{Henderson-Diggle-Dobson-2000}
%
\begin{barticle}[pbm]
\bauthor{\bsnm{Henderson},~\bfnm{R.}\binits{R.}},
\bauthor{\bsnm{Diggle},~\bfnm{P.}\binits{P.}} \AND
\bauthor{\bsnm{Dobson},~\bfnm{A.}\binits{A.}}
(\byear{2000}).
\btitle{Joint modelling of longitudinal measurements and event time data}.
\bjournal{Biostatistics}
\bvolume{1}
\bpages{465--480}.
\bid{doi={10.1093/biostatistics/1.4.465}, issn={1465-4644}, pii={1/4/465},
pmid={12933568}}
\bptok{imsref}%
\end{barticle}
%
\endbibitem

\bibitem[\protect\citeauthoryear{Hern{\'{a}}n, Clayton and
Keiding}{2011}]{Hernan-Clayton-Keiding-2011}
%
\begin{barticle}[pbm]
\bauthor{\bsnm{Hern{\'{a}}n},~\bfnm{Miguel~A.}\binits{M.~A.}},
\bauthor{\bsnm{Clayton},~\bfnm{David}\binits{D.}} \AND
\bauthor{\bsnm{Keiding},~\bfnm{Niels}\binits{N.}}
(\byear{2011}).
\btitle{The Simpson's paradox unraveled}.
\bjournal{Int. J. Epidemiol.}
\bvolume{40}
\bpages{780--785}.
\bid{doi={10.1093/ije/dyr041}, issn={1464-3685}, pii={dyr041}, pmcid={3147074},
pmid={21454324}}
\bptok{imsref}%
\end{barticle}
%
\endbibitem

\bibitem[\protect\citeauthoryear{Hern{\'a}n and
Robins}{2006}]{Hernan-Robins-2006}
%
\begin{barticle}[auto:STB|2013/12/09|07:59:19]
\bauthor{\bsnm{Hern{\'a}n},~\bfnm{M.~A.}\binits{M.~A.}} \AND
\bauthor{\bsnm{Robins},~\bfnm{J.~M.}\binits{J.~M.}}
(\byear{2006}).
\btitle{Estimating causal effects from epidemiological data}.
\bjournal{J. Epidemiol. Community Health}
\bvolume{60}
\bpages{578--586}.
\bptok{imsref}%
\end{barticle}
%
\endbibitem

\bibitem[\protect\citeauthoryear{Hobcraft, Menken and
Preston}{1982}]{Hobcraft-etal.-1982}
%
\begin{barticle}[pbm]
\bauthor{\bsnm{Hobcraft},~\bfnm{J.}\binits{J.}},
\bauthor{\bsnm{Menken},~\bfnm{J.}\binits{J.}} \AND
\bauthor{\bsnm{Preston},~\bfnm{S.}\binits{S.}}
(\byear{1982}).
\btitle{Age, period, and cohort effects in demography: A review}.
\bjournal{Popul. Index}
\bvolume{48}
\bpages{4--43}.
\bid{issn={0032-4701}, pmid={12338741}}
\bptok{imsref}%
\end{barticle}
%
\endbibitem

\bibitem[\protect\citeauthoryear{Hoem}{1987}]{Hoem-1987}
%
\begin{barticle}[auto:STB|2013/12/09|07:59:19]
\bauthor{\bsnm{Hoem},~\bfnm{J.}\binits{J.}}
(\byear{1987}).
\btitle{Statistical analysis of a multiplicative model and its
application to
the standardization of vital rates---A review}.
\bjournal{Int. Stat. Rev.}
\bvolume{55}
\bpages{119--152}.
\bptok{imsref}%
\end{barticle}
%
\endbibitem

\bibitem[\protect\citeauthoryear{Holford}{1980}]{Holford-1980}
%
\begin{barticle}[pbm]
\bauthor{\bsnm{Holford},~\bfnm{T.~R.}\binits{T.~R.}}
(\byear{1980}).
\btitle{The analysis of rates and of survivorship using log-linear models}.
\bjournal{Biometrics}
\bvolume{36}
\bpages{299--305}.
\bid{issn={0006-341X}, pmid={7407317}}
\bptok{imsref}%
\end{barticle}
%
\endbibitem

\bibitem[\protect\citeauthoryear{Hrobjartsson, G{\o}tzsche and
Gluud}{1998}]{Hrobjartsson-etal.-1998}
%
\begin{barticle}[auto:STB|2013/12/09|07:59:19]
\bauthor{\bsnm{Hrobjartsson},~\bfnm{A.}\binits{A.}},
\bauthor{\bsnm{G{\o}tzsche},~\bfnm{P.~C.}\binits{P.~C.}} \AND
\bauthor{\bsnm{Gluud},~\bfnm{C.}\binits{C.}}
(\byear{1998}).
\btitle{The controlled clinical trial turns 100 years: Fibiger's trial
of serum
treatment of diphtheria}.
\bjournal{Br. Med. J.}
\bvolume{317}
\bpages{1243--1245}.
\bptok{imsref}%
\end{barticle}
%
\endbibitem

\bibitem[\protect\citeauthoryear{Kalbfleisch and Prentice}{1980}]{kalbpren}
%
\begin{bbook}[mr]
\bauthor{\bsnm{Kalbfleisch},~\bfnm{John~D.}\binits{J.~D.}} \AND
\bauthor{\bsnm{Prentice},~\bfnm{Ross~L.}\binits{R.~L.}}
(\byear{1980}).
\btitle{The Statistical Analysis of Failure Time Data}.
\bpublisher{Wiley}, \blocation{New York}.
\bid{mr={0570114}}
\bptok{imsref}%
\end{bbook}
%
\endbibitem

\bibitem[\protect\citeauthoryear{Kalton}{1968}]{Kalton-1968}
%
\begin{barticle}[mr]
\bauthor{\bsnm{Kalton},~\bfnm{G.}\binits{G.}}
(\byear{1968}).
\btitle{Standardization: {A} technique to control for extraneous variables}.
\bjournal{Appl. Statist.}
\bvolume{17}
\bpages{118--136}.
\bid{mr={0234599}}
\bptok{imsref}%
\end{barticle}
%
\endbibitem

\bibitem[\protect\citeauthoryear{Keiding}{1987}]{Keiding-1987}
%
\begin{barticle}[mr]
\bauthor{\bsnm{Keiding},~\bfnm{Niels}\binits{N.}}
(\byear{1987}).
\btitle{The method of expected number of deaths, 1786--1886--1986}.
\bjournal{Internat. Statist. Rev.}
\bvolume{55}
\bpages{1--20}.
\bid{doi={10.2307/1403267}, issn={0306-7734}, mr={0962938}}
\bptok{imsref}%
\end{barticle}
%
\endbibitem

\bibitem[\protect\citeauthoryear{Keiding}{2011}]{Keiding-2011}
%
\begin{barticle}[mr]
\bauthor{\bsnm{Keiding},~\bfnm{Niels}\binits{N.}}
(\byear{2011}).
\btitle{Age-period-cohort analysis in the 1870s: Diagrams, stereograms,
and the
basic differential equation}.
\bjournal{Canad. J. Statist.}
\bvolume{39}
\bpages{405--420}.
\bid{issn={0319-5724}, mr={2842421}}
\bptok{imsref}%
\end{barticle}
%
\endbibitem

\bibitem[\protect\citeauthoryear{Kendall and
Lazarsfeld}{1950}]{Kendall-Lazarsfeld-1950}
%
\begin{bincollection}[auto:STB|2013/12/09|07:59:19]
\bauthor{\bsnm{Kendall},~\bfnm{P.~L.}\binits{P.~L.}} \AND
\bauthor{\bsnm{Lazarsfeld},~\bfnm{P.~F.}\binits{P.~F.}}
(\byear{1950}).
\btitle{Problems of survey analysis}.
In \bbooktitle{Continuities in Social Research: Studies in the Scope
and Method
of ``The American Soldier''}
(\beditor{\bfnm{R.}\binits{R.}~\bsnm{Merton}} \AND
\beditor{\bfnm{P.}\binits{P.}~\bsnm{Lazarsfeld}}, eds.)
\bpages{133--196}.
\bpublisher{Free Press}, \blocation{Glencoe, IL}.
\bptok{imsref}%
\end{bincollection}
%
\endbibitem

\bibitem[\protect\citeauthoryear{Kermack, Mc{K}endrick and Mc{K}inlay}{1934a}]{Kermack-McKendrick-McKinlay-1934a}
%
\begin{barticle}[auto:STB|2013/12/09|07:59:19]
\bauthor{\bsnm{Kermack},~\bfnm{W.~O.}\binits{W.~O.}},
\bauthor{\bsnm{Mc{K}endrick},~\bfnm{A.~G.}\binits{A.~G.}} \AND
\bauthor{\bsnm{Mc{K}inlay},~\bfnm{P.~L.}\binits{P.~L.}}
(\byear{1934}a).
\btitle{Death-rates in Great Britain and Sweden---Some general
regularities and
their significance}.
\bjournal{Lancet}
\bvolume{1}
\bpages{698--703}.
\bptok{imsref}%
\end{barticle}
%
\endbibitem

\bibitem[\protect\citeauthoryear{Kermack, Mc{K}endrick and
Mc{K}inlay}{1934b}]{Kermack-McKendrick-McKinlay-1934b}
%
\begin{barticle}[auto:STB|2013/12/09|07:59:19]
\bauthor{\bsnm{Kermack},~\bfnm{W.~O.}\binits{W.~O.}},
\bauthor{\bsnm{Mc{K}endrick},~\bfnm{A.~G.}\binits{A.~G.}} \AND
\bauthor{\bsnm{Mc{K}inlay},~\bfnm{P.~L.}\binits{P.~L.}}
(\byear{1934}b).
\btitle{Death-rates in Great Britain and Sweden: Expression of specific
mortality rates as products of two factors, and some consequences thereof}.
\bjournal{J. Hyg.}
\bvolume{34}
\bpages{433--457}.
\bptok{imsref}%
\end{barticle}
%
\endbibitem

\bibitem[\protect\citeauthoryear{Kilpatrick}{1962}]{Kilpatrick-1962}
%
\begin{barticle}[auto:STB|2013/12/09|07:59:19]
\bauthor{\bsnm{Kilpatrick},~\bfnm{S.~J.}\binits{S.~J.}}
(\byear{1962}).
\btitle{Occupational mortality indexes}.
\bjournal{Popul. Stud.}
\bvolume{16}
\bpages{175--187}.
\bptok{imsref}%
\end{barticle}
%
\endbibitem

\bibitem[\protect\citeauthoryear{Kitagawa}{1955}]{Kitagawa-1955}
%
\begin{barticle}[auto:STB|2013/12/09|07:59:19]
\bauthor{\bsnm{Kitagawa},~\bfnm{E.~M.}\binits{E.~M.}}
(\byear{1955}).
\btitle{Components of a difference between 2 rates}.
\bjournal{J. Amer. Statist. Assoc.}
\bvolume{50}
\bpages{1168--1194}.
\bptok{imsref}%
\end{barticle}
%
\endbibitem

\bibitem[\protect\citeauthoryear{Kitagawa}{1964}]{Kitagawa-1964}
%
\begin{barticle}[auto:STB|2013/12/09|07:59:19]
\bauthor{\bsnm{Kitagawa},~\bfnm{E.~M.}\binits{E.~M.}}
(\byear{1964}).
\btitle{Standardized comparisons in population-research}.
\bjournal{Demography}
\bvolume{1}
\bpages{296--315}.
\bptok{imsref}%
\end{barticle}
%
\endbibitem

\bibitem[\protect\citeauthoryear{Kitagawa}{1966}]{Kitagawa-1966}
%
\begin{barticle}[pbm]
\bauthor{\bsnm{Kitagawa},~\bfnm{E.~M.}\binits{E.~M.}}
(\byear{1966}).
\btitle{Theoretical considerations in the selection of a mortality index,
and some empirical comparisons}.
\bjournal{Hum. Biol.}
\bvolume{38}
\bpages{293--308}.
\bid{issn={0018-7143}, pmid={5977533}}
\bptok{imsref}%
\end{barticle}
%
\endbibitem

\bibitem[\protect\citeauthoryear{K{\"o}r{\"o}si}{1892--1893}]{Korosi-1892-1893}
%
\begin{barticle}[auto:STB|2013/12/09|07:59:19]
\bauthor{\bsnm{K{\"o}r{\"o}si},~\bfnm{J.}\binits{J.}}
(\byear{1892--1893}).
\btitle{Report of an international mortality standard, or mortality index}.
\bjournal{Pub. Amer. Statist. Assoc.}
\bvolume{3}
\bpages{450--462}.
\bptok{imsref}%
\end{barticle}
%
\endbibitem

\bibitem[\protect\citeauthoryear{Kurth et~al.}{2006}]{Kurth-etal.-2006}
%
\begin{barticle}[pbm]
\bauthor{\bsnm{Kurth},~\bfnm{Tobias}\binits{T.}},
\bauthor{\bsnm{Walker},~\bfnm{Alexander~M.}\binits{A.~M.}},
\bauthor{\bsnm{Glynn},~\bfnm{Robert~J.}\binits{R.~J.}},
\bauthor{\bsnm{Chan},~\bfnm{K.~Arnold}\binits{K.~A.}},
\bauthor{\bsnm{Gaziano},~\bfnm{J.~Michael}\binits{J.~M.}},
\bauthor{\bsnm{Berger},~\bfnm{Klaus}\binits{K.}} \AND
\bauthor{\bsnm{Robins},~\bfnm{James~M.}\binits{J.~M.}}
(\byear{2006}).
\btitle{Results of multivariable logistic regression, propensity matching,
propensity adjustment, and propensity-based weighting under conditions of
nonuniform effect}.
\bjournal{Am. J. Epidemiol.}
\bvolume{163}
\bpages{262--270}.
\bid{doi={10.1093/aje/kwj047}, issn={0002-9262}, pii={kwj047}, pmid={16371515}}
\bptok{imsref}%
\end{barticle}
%
\endbibitem

\bibitem[\protect\citeauthoryear{Lane and Nelder}{1982}]{Lane-Nelder-1982}
%
\begin{barticle}[pbm]
\bauthor{\bsnm{Lane},~\bfnm{P.~W.}\binits{P.~W.}} \AND
\bauthor{\bsnm{Nelder},~\bfnm{J.~A.}\binits{J.~A.}}
(\byear{1982}).
\btitle{Analysis of covariance and standardization as instances of prediction}.
\bjournal{Biometrics}
\bvolume{38}
\bpages{613--621}.
\bid{issn={0006-341X}, pmid={7171691}}
\bptok{imsref}%
\end{barticle}
%
\endbibitem

\bibitem[\protect\citeauthoryear{Lexis}{1876}]{Lexis-1876}
%
\begin{barticle}[auto:STB|2013/12/09|07:59:19]
\bauthor{\bsnm{Lexis},~\bfnm{W.}\binits{W.}}
(\byear{1876}).
\btitle{Das Geschlechtsverh\"altnis der Geborenen und die
Wahrscheinlichkeitsrechnung}.
\bjournal{Jahrb\"ucher f\"ur National\"okonomie und Statistik}
\bvolume{27}
\bpages{209--245}.
\bptok{imsref}%
\end{barticle}
%
\endbibitem

\bibitem[\protect\citeauthoryear{Liao}{1989}]{Liao-1989}
%
\begin{barticle}[pbm]
\bauthor{\bsnm{Liao},~\bfnm{T.~F.}\binits{T.~F.}}
(\byear{1989}).
\btitle{A flexible approach for the decomposition of rate differences}.
\bjournal{Demography}
\bvolume{26}
\bpages{717--726}.
\bid{issn={0070-3370}, pmid={2583328}}
\bptok{imsref}%
\end{barticle}
%
\endbibitem

\bibitem[\protect\citeauthoryear{MacKenzie}{1978}]{MacKenzie-1978}
%
\begin{barticle}[pbm]
\bauthor{\bsnm{MacKenzie},~\bfnm{D.}\binits{D.}}
(\byear{1978}).
\btitle{Statistical theory and social interests: A~case-study}.
\bjournal{Soc. Stud. Sci.}
\bvolume{8}
\bpages{35--83}.
\bid{issn={0306-3127}, pmid={11615698}}
\bptok{imsref}%
\end{barticle}
%
\endbibitem

\bibitem[\protect\citeauthoryear{MacKenzie}{1981}]{MacKenzie-1981}
%
\begin{bbook}[mr]
\bauthor{\bsnm{MacKenzie},~\bfnm{Donald~A.}\binits{D.~A.}}
(\byear{1981}).
\btitle{Statistics in {B}ritain: 1865--1930. The Social Construction of
Scientific Knowledge}.
\bpublisher{Edinburgh Univ. Press}, \blocation{Edinburgh}.
\bid{mr={0642200}}
\bptok{imsref}%
\end{bbook}
%
\endbibitem

\bibitem[\protect\citeauthoryear{MacKinnon}{2008}]{MacKinnon-2008}
%
\begin{bmisc}[auto:STB|2013/12/09|07:59:19]
\bauthor{\bsnm{MacKinnon},~\bfnm{D.}\binits{D.}}
(\byear{2008}).
\bhowpublished{\textit{Introduction to Statistical Mediation Analysis}.
Lawrence Erbaum Associates, New York.}
\bptok{imsref}%
\end{bmisc}
%
\endbibitem

\bibitem[\protect\citeauthoryear{Mantel}{1966}]{mantel:66}
%
\begin{barticle}[auto:STB|2013/12/09|07:59:19]
\bauthor{\bsnm{Mantel},~\bfnm{N.}\binits{N.}}
(\byear{1966}).
\btitle{Evaluation of survival data and two new rank order statistics arising
in its consideration}.
\bjournal{Cancer Chemotherapy Reports}
\bvolume{50}
\bpages{163--170}.
\bptok{imsref}%
\end{barticle}
%
\endbibitem

\bibitem[\protect\citeauthoryear{Mantel and
Haenszel}{1959}]{mantel:haenszel:1959}
%
\begin{barticle}[auto:STB|2013/12/09|07:59:19]
\bauthor{\bsnm{Mantel},~\bfnm{N.}\binits{N.}} \AND
\bauthor{\bsnm{Haenszel},~\bfnm{W.}\binits{W.}}
(\byear{1959}).
\btitle{Statistical aspects of the analysis of data from retrospective studies
of disease}.
\bjournal{J. Natl. Cancer. Inst.}
\bvolume{22}
\bpages{719--748}.
\bptok{imsref}%
\end{barticle}
%
\endbibitem

\bibitem[\protect\citeauthoryear{Mantel and Stark}{1968}]{Mantel-Stark-1968}
%
\begin{barticle}[pbm]
\bauthor{\bsnm{Mantel},~\bfnm{N.}\binits{N.}} \AND
\bauthor{\bsnm{Stark},~\bfnm{C.~R.}\binits{C.~R.}}
(\byear{1968}).
\btitle{Computation of indirect-adjusted rates in the presence of confounding}.
\bjournal{Biometrics}
\bvolume{24}
\bpages{997--1005}.
\bid{issn={0006-341X}, pmid={4236976}}
\bptok{imsref}%
\end{barticle}
%
\endbibitem

\bibitem[\protect\citeauthoryear
{Mc{K}endrick}{1925--1926}]{McKendrick-1925-19%
26}
%
\begin{barticle}[auto:STB|2013/12/09|07:59:19]
\bauthor{\bsnm{Mc{K}endrick},~\bfnm{A.~G.}\binits{A.~G.}}
(\byear{1925--1926}).
\btitle{Applications of mathematics to medical problems}.
\bjournal{Proc. Edinb. Math. Soc. (2)}
\bvolume{43--44}
\bpages{98--130}.
\bptok{imsref}%
\end{barticle}
%
\endbibitem

\bibitem[\protect\citeauthoryear{Mc{K}inlay}{1975}]{McKinlay-1975}
%
\begin{barticle}[auto:STB|2013/12/09|07:59:19]
\bauthor{\bsnm{Mc{K}inlay},~\bfnm{S.~M.}\binits{S.~M.}}
(\byear{1975}).
\btitle{Design and analysis of observational study---Review}.
\bjournal{J. Amer. Statist. Assoc.}
\bvolume{70}
\bpages{503--520}.
\bptok{imsref}%
\end{barticle}
%
\endbibitem

\bibitem[\protect\citeauthoryear{Miettinen}{1972a}]{miettinen:1972a}
%
\begin{barticle}[auto:STB|2013/12/09|07:59:19]
\bauthor{\bsnm{Miettinen},~\bfnm{O.}\binits{O.}}
(\byear{1972}a).
\btitle{Standardization of risk ratios}.
\bjournal{Am. J. Epidemiol.}
\bvolume{96}
\bpages{383--388}.
\bptok{imsref}%
\end{barticle}
%
\endbibitem

\bibitem[\protect\citeauthoryear{Miettinen}{1972b}]{Miettinen-1972}
%
\begin{barticle}[auto:STB|2013/12/09|07:59:19]
\bauthor{\bsnm{Miettinen},~\bfnm{O.~S.}\binits{O.~S.}}
(\byear{1972}b).
\btitle{Components of the crude risk ratio}.
\bjournal{Am. J. Epidemiol.}
\bvolume{96}
\bpages{168--172}.
\bptok{imsref}%
\end{barticle}
%
\endbibitem

\bibitem[\protect\citeauthoryear{Miettinen}{1976a}]{miettinen:76}
%
\begin{barticle}[auto:STB|2013/12/09|07:59:19]
\bauthor{\bsnm{Miettinen},~\bfnm{O.}\binits{O.}}
(\byear{1976}a).
\btitle{Estimability and estimation in case-referent studies}.
\bjournal{Am. J. Epidemiol.}
\bvolume{103}
\bpages{230--235}.
\bptok{imsref}%
\end{barticle}
%
\endbibitem

\bibitem[\protect\citeauthoryear{Miettinen}{1976b}]{Miettinen-1976}
%
\begin{barticle}[auto:STB|2013/12/09|07:59:19]
\bauthor{\bsnm{Miettinen},~\bfnm{O.~S.}\binits{O.~S.}}
(\byear{1976}b).
\btitle{Stratification by a multivariate confounder score}.
\bjournal{Am. J. Epidemiol.}
\bvolume{104}
\bpages{609--620}.
\bptok{imsref}%
\end{barticle}
%
\endbibitem

\bibitem[\protect\citeauthoryear{Miettinen}{1985}]{Miettinen-1985}
%
\begin{bbook}[auto:STB|2013/12/09|07:59:19]
\bauthor{\bsnm{Miettinen},~\bfnm{O.~S.}\binits{O.~S.}}
(\byear{1985}).
\btitle{Theoretical Epidemiology}.
\bpublisher{Wiley}, \blocation{New York}.
\bptok{imsref}%
\end{bbook}
%
\endbibitem

\bibitem[\protect\citeauthoryear{Neison}{1844}]{Neison-1844}
%
\begin{barticle}[auto:STB|2013/12/09|07:59:19]
\bauthor{\bsnm{Neison},~\bfnm{F.}\binits{F.}}
(\byear{1844}).
\btitle{On a method recently proposed for conducting inquiries into the
comparative sanatory condition of various districts, with illustrations,
derived from numerous places in Great Britain at the period of the last
census}.
\bjournal{J. Stat. Soc. London}
\bvolume{7}
\bpages{40--68}.
\bptok{imsref}%
\end{barticle}
%
\endbibitem

\bibitem[\protect\citeauthoryear{Neison}{1851}]{Neison-1851}
%
\begin{barticle}[auto:STB|2013/12/09|07:59:19]
\bauthor{\bsnm{Neison},~\bfnm{F.~G.~P.}\binits{F.~G.~P.}}
(\byear{1851}).
\btitle{On the rate of mortality among persons of intemperate habits}.
\bjournal{J. Stat. Soc. London}
\bvolume{14}
\bpages{200--219}.
\bptok{imsref}%
\end{barticle}
%
\endbibitem

\bibitem[\protect\citeauthoryear{Nelder and
Wedderburn}{1972}]{nelder:wedderburn:72}
%
\begin{barticle}[auto:STB|2013/12/09|07:59:19]
\bauthor{\bsnm{Nelder},~\bfnm{J.}\binits{J.}} \AND
\bauthor{\bsnm{Wedderburn},~\bfnm{R.}\binits{R.}}
(\byear{1972}).
\btitle{Generalized linear models}.
\bjournal{J. Roy. Statist. Soc. Ser. A}
\bvolume{135}
\bpages{370--384}.
\bptok{imsref}%
\end{barticle}
%
\endbibitem

\bibitem[\protect\citeauthoryear{Nelson}{1972}]{nelson:72}
%
\begin{barticle}[auto:STB|2013/12/09|07:59:19]
\bauthor{\bsnm{Nelson},~\bfnm{W.}\binits{W.}}
(\byear{1972}).
\btitle{Theory and applications of hazard plotting for censored failure data}.
\bjournal{Technometrics}
\bvolume{14}
\bpages{945--966}.
\bptok{imsref}%
\end{barticle}
%
\endbibitem

\bibitem[\protect\citeauthoryear{Newsholme and
Stevenson}{1906}]{Newsholme-Stevenson-1906}
%
\begin{barticle}[auto:STB|2013/12/09|07:59:19]
\bauthor{\bsnm{Newsholme},~\bfnm{A.}\binits{A.}} \AND
\bauthor{\bsnm{Stevenson},~\bfnm{T.~H.~C.}\binits{T.~H.~C.}}
(\byear{1906}).
\btitle{The decline of human fertility in the United Kingdom and other
countries as shown by corrected birth-rates}.
\bjournal{J. Roy. Stat. Soc.}
\bvolume{69}
\bpages{34--87}.
\bptok{imsref}%
\end{barticle}
%
\endbibitem

\bibitem[\protect\citeauthoryear{Ogle}{1892}]{Ogle-1892}
%
\begin{barticle}[auto:STB|2013/12/09|07:59:19]
\bauthor{\bsnm{Ogle},~\bfnm{W.}\binits{W.}}
(\byear{1892}).
\btitle{Proposal for the establishment and international use of a standard
population, with fixed sex and age distribution, in the calculation and
comparison of marriage, birth, and death rates}.
\bjournal{Bull. Int. Stat. Inst.}
\bvolume{6}
\bpages{83--85}.
\bptok{imsref}%
\end{barticle}
%
\endbibitem

\bibitem[\protect\citeauthoryear{Osborn}{1975}]{Osborn-1975}
%
\begin{barticle}[auto:STB|2013/12/09|07:59:19]
\bauthor{\bsnm{Osborn},~\bfnm{J.}\binits{J.}}
(\byear{1975}).
\btitle{Multiplicative model for analysis of vital statistics rates}.
\bjournal{Appl. Stat.}
\bvolume{24}
\bpages{75--84}.
\bptok{imsref}%
\end{barticle}
%
\endbibitem



\bibitem[\protect\citeauthoryear{Pearl}{2001}]{Pearl-2001}
%
\begin{bincollection}[auto:STB|2013/12/09|07:59:19]
\bauthor{\bsnm{Pearl},~\bfnm{J.}\binits{J.}}
(\byear{2001}).
\btitle{Direct and indirect effects}.
In \bbooktitle{Proceedings of the Seventeenth Conference on Uncertainty in
Artificial Intelligence}
\bpages{411--420}.
\bpublisher{Morgan Kaufman}, \blocation{San Francisco, CA}.
\bptok{imsref}%
\end{bincollection}
%
\endbibitem

\bibitem[\protect\citeauthoryear{Pearl}{2009}]{Pearl-2009}
%
\begin{bbook}[mr]
\bauthor{\bsnm{Pearl},~\bfnm{Judea}\binits{J.}}
(\byear{2009}).
\btitle{Causality: Models, Reasoning, and Inference},
\bedition{2nd} ed.
\bpublisher{Cambridge Univ. Press}, \blocation{Cambridge}.
\bid{mr={2548166}}
\bptok{imsref}%
\end{bbook}
%
\endbibitem

\bibitem[\protect\citeauthoryear{Pearl and
Barenboim}{2012}]{Pearl-Barenboim-2012}
%
\begin{bmisc}[auto:STB|2013/12/09|07:59:19]
\bauthor{\bsnm{Pearl},~\bfnm{J.}\binits{J.}} \AND
\bauthor{\bsnm{Barenboim},~\bfnm{E.}\binits{E.}}
(\byear{2012}).
\bhowpublished{Transportability across studies: A formal approach. Technical
report, Computer Science Dept., Univ. California, Los Angeles}.
\bptok{imsref}%
\end{bmisc}
%
\endbibitem

\bibitem[\protect\citeauthoryear{Pearson}{1900}]{Pearson-1900}
%
\begin{barticle}[auto:STB|2013/12/09|07:59:19]
\bauthor{\bsnm{Pearson},~\bfnm{K.}\binits{K.}}
(\byear{1900}).
\btitle{On the correlation of characters not quantitatively measurable}.
\bjournal{Philos. Trans. R. Soc. Lond. Ser. A Math. Phys. Eng. Sci.}
\bvolume{195}
\bpages{1--47}.
\bptok{imsref}%
\end{barticle}
%
\endbibitem

\bibitem[\protect\citeauthoryear{Pearson}{1910}]{Pearson-1910}
%
\begin{bbook}[auto:STB|2013/12/09|07:59:19]
\bauthor{\bsnm{Pearson},~\bfnm{K.}\binits{K.}}
(\byear{1910}).
\btitle{The Grammar of Science},
\bedition{3rd} ed.
\bpublisher{Black}, \blocation{Edinburgh}.
\bptok{imsref}%
\end{bbook}
%
\endbibitem

\bibitem[\protect\citeauthoryear{Pearson, Lee and
Bramley-{M}oore}{1899}]{Pearson-Lee-Bramley-Moore-1899}
%
\begin{barticle}[auto:STB|2013/12/09|07:59:19]
\bauthor{\bsnm{Pearson},~\bfnm{K.}\binits{K.}},
\bauthor{\bsnm{Lee},~\bfnm{A.}\binits{A.}} \AND
\bauthor{\bsnm{Bramley-{M}oore},~\bfnm{L.}\binits{L.}}
(\byear{1899}).
\btitle{Genetic (reproductive) selection: Inheritance of fertility in
man and
of fecundity in thoroughbred racehorses}.
\bjournal{Philos. Trans. R. Soc. Lond. Ser. A Math. Phys. Eng. Sci.}
\bvolume{192}
\bpages{534--539}.
\bptok{imsref}%
\end{barticle}
%
\endbibitem

\bibitem[\protect\citeauthoryear{Pearson and
Tocher}{1915}]{Pearson-Tocher-1916}
%
\begin{barticle}[auto:STB|2013/12/09|07:59:19]
\bauthor{\bsnm{Pearson},~\bfnm{K.}\binits{K.}} \AND
\bauthor{\bsnm{Tocher},~\bfnm{J.~F.}\binits{J.~F.}}
(\byear{1915}).
\btitle{On criteria for the existence of differential deathrates}.
\bjournal{Biometrika}
\bvolume{11}
\bpages{159--184}.
\bptok{imsref}%
\end{barticle}
%
\endbibitem

\bibitem[\protect\citeauthoryear{Peters}{1941}]{Peters-1941}
%
\begin{barticle}[auto:STB|2013/12/09|07:59:19]
\bauthor{\bsnm{Peters},~\bfnm{C.~C.}\binits{C.~C.}}
(\byear{1941}).
\btitle{A method of matching groups for experiment with no loss of population}.
\bjournal{J. Educ. Res.}
\bvolume{34}
\bpages{606--612}.
\bptok{imsref}%
\end{barticle}
%
\endbibitem

\bibitem[\protect\citeauthoryear{Petersen, Sinisi and van~der
Laan}{2006}]{Petersen-etal.-2006}
%
\begin{barticle}[pbm]
\bauthor{\bsnm{Petersen},~\bfnm{Maya~L.}\binits{M.~L.}},
\bauthor{\bsnm{Sinisi},~\bfnm{Sandra~E.}\binits{S.~E.}} \AND
\bauthor{\bparticle{van~der} \bsnm{Laan},~\bfnm{Mark~J.}\binits{M.~J.}}
(\byear{2006}).
\btitle{Estimation of direct causal effects}.
\bjournal{Epidemiology}
\bvolume{17}
\bpages{276--284}.
\bid{doi={10.1097/01.ede.0000208475.99429.2d}, issn={1044-3983},
pii={00001648-200605000-00012}, pmid={16617276}}
\bptok{imsref}%
\end{barticle}
%
\endbibitem

\bibitem[\protect\citeauthoryear{Pike, Anderson and
Day}{1979}]{Pike-etal.-1979}
%
\begin{barticle}[auto:STB|2013/12/09|07:59:19]
\bauthor{\bsnm{Pike},~\bfnm{M.}\binits{M.}},
\bauthor{\bsnm{Anderson},~\bfnm{J.}\binits{J.}} \AND
\bauthor{\bsnm{Day},~\bfnm{N.}\binits{N.}}
(\byear{1979}).
\btitle{Some insights into Miettinen's multivariate confounder score approach
to case-control study analysis}.
\bjournal{Epidemiol. Comm. Health}
\bvolume{33}
\bpages{104--106}.
\bptok{imsref}%
\end{barticle}
%
\endbibitem

\bibitem[\protect\citeauthoryear{Powers and Xie}{2008}]{Powers-Xie-2008}
%
\begin{bbook}[auto:STB|2013/12/09|07:59:19]
\bauthor{\bsnm{Powers},~\bfnm{D.~A.}\binits{D.~A.}} \AND
\bauthor{\bsnm{Xie},~\bfnm{Y.}\binits{Y.}}
(\byear{2008}).
\btitle{Statistical Methods for Categorical Data Analysis},
\bedition{2nd} ed.
\bpublisher{Emerald},
\blocation{Bingley, UK}.
\bptok{imsref}%
\end{bbook}
%
\endbibitem

\bibitem[\protect\citeauthoryear{Powers and Yun}{2009}]{Powers-Yun-2009}
%
\begin{barticle}[auto:STB|2013/12/09|07:59:19]
\bauthor{\bsnm{Powers},~\bfnm{D.~A.}\binits{D.~A.}} \AND
\bauthor{\bsnm{Yun},~\bfnm{M.~S.}\binits{M.~S.}}
(\byear{2009}).
\btitle{Multivariate decomposition for hazard rate models}.
\bjournal{Sociol. Methodol.}
\bvolume{39}
\bpages{233--263}.
\bptok{imsref}%
\end{barticle}
%
\endbibitem

\bibitem[\protect\citeauthoryear{Preston, Heuveline and
Guillot}{2001}]{Preston-etal.-2001}
%
\begin{bbook}[auto:STB|2013/12/09|07:59:19]
\bauthor{\bsnm{Preston},~\bfnm{S.~H.}\binits{S.~H.}},
\bauthor{\bsnm{Heuveline},~\bfnm{P.}\binits{P.}} \AND
\bauthor{\bsnm{Guillot},~\bfnm{M.}\binits{M.}}
(\byear{2001}).
\btitle{Demography}.
\bpublisher{Blackwell}, \blocation{Oxford}.
\bptok{imsref}%
\end{bbook}
%
\endbibitem

\bibitem[\protect\citeauthoryear{Robins}{1986}]{Robins-1986}
%
\begin{barticle}[mr]
\bauthor{\bsnm{Robins},~\bfnm{James}\binits{J.}}
(\byear{1986}).
\btitle{A new approach to causal inference in mortality studies with a
sustained exposure period---application to control of the healthy worker
survivor effect}.
\bjournal{Math. Modelling}
\bvolume{7}
\bpages{1393--1512}.
\bid{doi={10.1016/0270-0255(86)90088-6}, issn={0270-0255}, mr={0877758}}
\bptok{imsref}%
\end{barticle}
%
\endbibitem



\bibitem[\protect\citeauthoryear{Robins et~al.}{1992}]{Robins-etal.-1992}
%
\begin{barticle}[auto:STB|2013/12/09|07:59:19]
\bauthor{\bsnm{Robins},~\bfnm{J.}\binits{J.}},
\bauthor{\bsnm{Blevins},~\bfnm{D.}\binits{D.}},
\bauthor{\bsnm{Ritter},~\bfnm{G.}\binits{G.}} \AND
\bauthor{\bsnm{Wulfsohn},~\bfnm{M.}\binits{M.}}
(\byear{1992}).
\btitle{G-estimation of the effect of prophylaxis therapy for pneumocystis
carinii pneumonia on the survival of AIDS patients}.
\bjournal{Epidemiology}
\bvolume{14}
\bpages{79--97}.
\bptok{imsref}%
\end{barticle}
%
\endbibitem


\bibitem[\protect\citeauthoryear{Robins and
Hern{\'a}n}{2009}]{Robins-Hernan-2009}
%
\begin{bincollection}[mr]
\bauthor{\bsnm{Robins},~\bfnm{James~M.}\binits{J.~M.}} \AND
\bauthor{\bsnm{Hern{\'a}n},~\bfnm{Miguel~A.}\binits{M.~A.}}
(\byear{2009}).
\btitle{Estimation of the causal effects of time-varying exposures}.
In \bbooktitle{Longitudinal Data Analysis}
(\beditor{\bfnm{G.}\binits{G.}~\bsnm{Fitzmaurice}},
\beditor{\bfnm{G.}\binits{G.}~\bsnm{Verbeke}},
\beditor{\bfnm{M.}\binits{M.}~\bsnm{Davidian}} \AND
\beditor{\bfnm{G.}\binits{G.}~\bsnm{Molenberghs}}, eds.)
\bpages{553--599}.
\bpublisher{CRC Press}, \blocation{Boca Raton, FL}.
\bid{mr={1500133}}
\bptok{imsref}%
\end{bincollection}
%
\endbibitem


\bibitem[\protect\citeauthoryear{Robins, Hern{\'{a}}n and
Brumback}{2000}]{Robins-etal.-2000}
%
\begin{barticle}[pbm]
\bauthor{\bsnm{Robins},~\bfnm{J.~M.}\binits{J.~M.}},
\bauthor{\bsnm{Hern{\'{a}}n},~\bfnm{M.~A.}\binits{M.~A.}} \AND
\bauthor{\bsnm{Brumback},~\bfnm{B.}\binits{B.}}
(\byear{2000}).
\btitle{Marginal structural models and causal inference in epidemiology}.
\bjournal{Epidemiology}
\bvolume{11}
\bpages{550--560}.
\bid{issn={1044-3983}, pmid={10955408}}
\bptok{imsref}%
\end{barticle}
%
\endbibitem


\bibitem[\protect\citeauthoryear{Robins and
Tsiatis}{1992}]{Robins-Tsiatis-1992}
%
\begin{barticle}[mr]
\bauthor{\bsnm{Robins},~\bfnm{James}\binits{J.}} \AND
\bauthor{\bsnm{Tsiatis},~\bfnm{Anastasios~A.}\binits{A.~A.}}
(\byear{1992}).
\btitle{Semiparametric estimation of an accelerated failure time model with
time-dependent covariates}.
\bjournal{Biometrika}
\bvolume{79}
\bpages{311--319}.
\bid{issn={0006-3444}, mr={1185133}}
\bptok{imsref}%
\end{barticle}
%
\endbibitem


\bibitem[\protect\citeauthoryear{Rosenbaum}{1984}]{Rosenbaum-1984}
%
\begin{barticle}[mr]
\bauthor{\bsnm{Rosenbaum},~\bfnm{Paul~R.}\binits{P.~R.}}
(\byear{1984}).
\btitle{Conditional permutation tests and the propensity score in observational
studies}.
\bjournal{J. Amer. Statist. Assoc.}
\bvolume{79}
\bpages{565--574}.
\bid{issn={0162-1459}, mr={0763575}}
\bptok{imsref}%
\end{barticle}
%
\endbibitem

\bibitem[\protect\citeauthoryear{Rosenbaum and
Rubin}{1983}]{Rosenbaum-Rubin-1983}
%
\begin{barticle}[mr]
\bauthor{\bsnm{Rosenbaum},~\bfnm{Paul~R.}\binits{P.~R.}} \AND
\bauthor{\bsnm{Rubin},~\bfnm{Donald~B.}\binits{D.~B.}}
(\byear{1983}).
\btitle{The central role of the propensity score in observational
studies for
causal effects}.
\bjournal{Biometrika}
\bvolume{70}
\bpages{41--55}.
\bid{doi={10.1093/biomet/70.1.41}, issn={0006-3444}, mr={0742974}}
\bptok{imsref}%
\end{barticle}
%
\endbibitem

\bibitem[\protect\citeauthoryear{Rosenberg}{1962}]{Rosenberg-1962}
%
\begin{barticle}[auto:STB|2013/12/09|07:59:19]
\bauthor{\bsnm{Rosenberg},~\bfnm{M.}\binits{M.}}
(\byear{1962}).
\btitle{Test factor standardization as a method of interpretation}.
\bjournal{Soc. Forces}
\bvolume{41}
\bpages{53--61}.
\bptok{imsref}%
\end{barticle}
%
\endbibitem

\bibitem[\protect\citeauthoryear{Rothman}{1986}]{Rothman-1986}
%
\begin{bbook}[auto:STB|2013/12/09|07:59:19]
\bauthor{\bsnm{Rothman},~\bfnm{K.}\binits{K.}}
(\byear{1986}).
\btitle{Modern Epidemiology}.
\bpublisher{Little, Brown and Company},
\blocation{Boston, MA}.
\bptok{imsref}%
\end{bbook}
%
\endbibitem

\bibitem[\protect\citeauthoryear{Rothman and
Greenland}{1998}]{Rothman-Greenland-1998}
%
\begin{bbook}[auto:STB|2013/12/09|07:59:19]
\bauthor{\bsnm{Rothman},~\bfnm{K.}\binits{K.}} \AND
\bauthor{\bsnm{Greenland},~\bfnm{S.}\binits{S.}}
(\byear{1998}).
\btitle{Modern Epidemiology},
\bedition{2nd} ed.
\bpublisher{Lippincott, Williams \& Wilkins},
\blocation{Philadelphia, PA}.
\bptok{imsref}%
\end{bbook}
%
\endbibitem

\bibitem[\protect\citeauthoryear{Rothman, Greenland and
Lash}{2008}]{Rothman-etal.-2008}
%
\begin{bbook}[auto:STB|2013/12/09|07:59:19]
\bauthor{\bsnm{Rothman},~\bfnm{K.~J.}\binits{K.~J.}},
\bauthor{\bsnm{Greenland},~\bfnm{S.}\binits{S.}} \AND
\bauthor{\bsnm{Lash},~\bfnm{T.~L.}\binits{T.~L.}}
(\byear{2008}).
\btitle{Modern Epidemiology},
\bedition{3rd} ed.
\bpublisher{Lippincott, Williams \& Wilkins},
\blocation{Philadelphia, PA}.
\bptok{imsref}%
\end{bbook}
%
\endbibitem

\bibitem[\protect\citeauthoryear{Sato and
Matsuyama}{2003}]{Sato-Matsuyama-2003}
%
\begin{barticle}[pbm]
\bauthor{\bsnm{Sato},~\bfnm{Tosiya}\binits{T.}} \AND
\bauthor{\bsnm{Matsuyama},~\bfnm{Yutaka}\binits{Y.}}
(\byear{2003}).
\btitle{Marginal structural models as a tool for standardization}.
\bjournal{Epidemiology}
\bvolume{14}
\bpages{680--686}.
\bid{doi={10.1097/01.EDE.0000081989.82616.7d}, issn={1044-3983},
pmid={14569183}}
\bptok{imsref}%
\end{barticle}
%
\endbibitem

\bibitem[\protect\citeauthoryear{Schweber}{2001}]{Schweber-2001}
%
\begin{barticle}[pbm]
\bauthor{\bsnm{Schweber},~\bfnm{L.}\binits{L.}}
(\byear{2001}).
\btitle{Manipulation and population statistics in nineteenth-century
France and
England}.
\bjournal{Soc. Res. (New York)}
\bvolume{68}
\bpages{547--582}.
\bid{issn={0037-783X}, pmid={18574894}}
\bptok{imsref}%
\end{barticle}
%
\endbibitem

\bibitem[\protect\citeauthoryear{Schweber}{2006}]{Schweber-2006}
%
\begin{bbook}[auto:STB|2013/12/09|07:59:19]
\bauthor{\bsnm{Schweber},~\bfnm{L.}\binits{L.}}
(\byear{2006}).
\btitle{Disciplining Statistics: Demography and Vital Statistics in
France and
England}.
\bpublisher{Duke Univ. Press}, \blocation{Durham}.
\bptok{imsref}%
\end{bbook}
%
\endbibitem

\bibitem[\protect\citeauthoryear{Scott}{1978}]{Scott-1978}
%
\begin{barticle}[auto:STB|2013/12/09|07:59:19]
\bauthor{\bsnm{Scott},~\bfnm{R.~C.}\binits{R.~C.}}
(\byear{1978}).
\btitle{Bias problem in smear-and-sweep analysis}.
\bjournal{J. Amer. Statist. Assoc.}
\bvolume{73}
\bpages{714--718}.
\bptok{imsref}%
\end{barticle}
%
\endbibitem

\bibitem[\protect\citeauthoryear{Searle, Speed and
Milliken}{1980}]{Searle-etal.-1980}
%
\begin{barticle}[mr]
\bauthor{\bsnm{Searle},~\bfnm{S.~R.}\binits{S.~R.}},
\bauthor{\bsnm{Speed},~\bfnm{F.~M.}\binits{F.~M.}} \AND
\bauthor{\bsnm{Milliken},~\bfnm{G.~A.}\binits{G.~A.}}
(\byear{1980}).
\btitle{Population marginal means in the linear model: An alternative
to least
squares means}.
\bjournal{Amer. Statist.}
\bvolume{34}
\bpages{216--221}.
\bid{doi={10.2307/2684063}, issn={0003-1305}, mr={0596242}}
\bptok{imsref}%
\end{barticle}
%
\endbibitem

\bibitem[\protect\citeauthoryear{Simpson}{1951}]{Simpson-1951}
%
\begin{barticle}[mr]
\bauthor{\bsnm{Simpson},~\bfnm{E.~H.}\binits{E.~H.}}
(\byear{1951}).
\btitle{The interpretation of interaction in contingency tables}.
\bjournal{J. R. Stat. Soc. Ser. B Stat. Methodol.}
\bvolume{13}
\bpages{238--241}.
\bid{issn={0035-9246}, mr={0051472}}
\bptok{imsref}%
\end{barticle}
%
\endbibitem

\bibitem[\protect\citeauthoryear{Sobel}{1996}]{Sobel-1996}
%
\begin{barticle}[auto:STB|2013/12/09|07:59:19]
\bauthor{\bsnm{Sobel},~\bfnm{M.~E.}\binits{M.~E.}}
(\byear{1996}).
\btitle{Clifford Collier Clogg, 1949--1995: A tribute to his life and work}.
\bjournal{Sociol. Methodol.}
\bvolume{26}
\bpages{1--38}.
\bptok{imsref}%
\end{barticle}
%
\endbibitem

\bibitem[\protect\citeauthoryear{Stigler}{1986}]{Stigler-1986}
%
\begin{bbook}[mr]
\bauthor{\bsnm{Stigler},~\bfnm{Stephen~M.}\binits{S.~M.}}
(\byear{1986}).
\btitle{The History of Statistics: The Measurement of Uncertainty
Before 1900}.
\bpublisher{The Belknap Press of Harvard Univ. Press}, \blocation{Cambridge,
MA}.
\bid{mr={0852410}}
\bptok{imsref}%
\end{bbook}
%
\endbibitem

\bibitem[\protect\citeauthoryear{Stone}{1970}]{Stone-1970}
%
\begin{barticle}[auto:STB|2013/12/09|07:59:19]
\bauthor{\bsnm{Stone},~\bfnm{M.}\binits{M.}}
(\byear{1970}).
\btitle{Review of National Halothane Study---Bunker, {J.~P}., Forrest, {W.~H.},
Mosteller, {F.} and Vandam, {L.~D.~D}}.
\bjournal{J. Amer. Statist. Assoc.}
\bvolume{65}
\bpages{1392--1396}.
\bptok{imsref}%
\end{barticle}
%
\endbibitem

\bibitem[\protect\citeauthoryear{Stouffer and
Tibbitts}{1933}]{Stouffer-Tibbitts-1933}
%
\begin{barticle}[auto:STB|2013/12/09|07:59:19]
\bauthor{\bsnm{Stouffer},~\bfnm{S.~A.}\binits{S.~A.}} \AND
\bauthor{\bsnm{Tibbitts},~\bfnm{C.}\binits{C.}}
(\byear{1933}).
\btitle{Tests of significance in applying Westergaard's method of expected
cases to sociological data}.
\bjournal{J. Amer. Statist. Assoc.}
\bvolume{28}
\bpages{293--302}.
\bptok{imsref}%
\end{barticle}
%
\endbibitem

\bibitem[\protect\citeauthoryear{Thiele}{1881}]{Thiele-1881}
%
\begin{barticle}[auto:STB|2013/12/09|07:59:19]
\bauthor{\bsnm{Thiele},~\bfnm{T.}\binits{T.}}
(\byear{1881}).
\btitle{H. Westergaards Mortalitets-{S}tatistik}.
\bjournal{National{\o}konomisk Tidsskrift}
\bvolume{18}
\bpages{219--227}.
\bptok{imsref}%
\end{barticle}
%
\endbibitem

\bibitem[\protect\citeauthoryear{Truett, Cornfield and
Kannel}{1967}]{truett:cornfield:kannel:1967}
%
\begin{barticle}[pbm]
\bauthor{\bsnm{Truett},~\bfnm{J.}\binits{J.}},
\bauthor{\bsnm{Cornfield},~\bfnm{J.}\binits{J.}} \AND
\bauthor{\bsnm{Kannel},~\bfnm{W.}\binits{W.}}
(\byear{1967}).
\btitle{A multivariate analysis of the risk of coronary heart disease in
Framingham}.
\bjournal{J. Chronic. Dis.}
\bvolume{20}
\bpages{511--524}.
\bid{issn={0021-9681}, pmid={6028270}}
\bptok{imsref}%
\end{barticle}
%
\endbibitem

\bibitem[\protect\citeauthoryear{Tukey}{1979}]{Tukey-1979}
%
\begin{barticle}[mr]
\bauthor{\bsnm{Tukey},~\bfnm{J.~W.}\binits{J.~W.}}
(\byear{1979}).
\btitle{Methodology, and the statisticians responsibility for both accuracy
 and relevance}.
\bjournal{J.~Amer. Statist. Assoc.}
\bvolume{74}
\bpages{786--793}.
\bptok{imsref}%
\end{barticle}
%
\endbibitem


\bibitem[\protect\citeauthoryear{Tukey}{1991}]{Tukey-1991}
%
\begin{barticle}[auto:STB|2013/12/09|07:59:19]
\bauthor{\bsnm{Tukey},~\bfnm{J.~W.}\binits{J.~W.}}
(\byear{1991}).
\btitle{Use of many covariates in clinical trials}.
\bjournal{Int. Stat. Rev.}
\bvolume{59}
\bpages{123--137}.
\bptok{imsref}%
\end{barticle}
%
\endbibitem

\bibitem[\protect\citeauthoryear{Turner}{1949}]{Turner-1949}
%
\begin{barticle}[auto:STB|2013/12/09|07:59:19]
\bauthor{\bsnm{Turner},~\bfnm{R.~H.}\binits{R.~H.}}
(\byear{1949}).
\btitle{The expected-cases method applied to the nonwhite male labor force}.
\bjournal{Am. J. Sociol.}
\bvolume{55}
\bpages{146--156}.
\bptok{imsref}%
\end{barticle}
%
\endbibitem

\bibitem[\protect\citeauthoryear{Vansteelandt and
Keiding}{2011}]{Vansteelandt-Keiding-2011}
%
\begin{barticle}[auto:STB|2013/12/09|07:59:19]
\bauthor{\bsnm{Vansteelandt},~\bfnm{S.}\binits{S.}} \AND
\bauthor{\bsnm{Keiding},~\bfnm{N.}\binits{N.}}
(\byear{2011}).
\btitle{Invited commentary: G-computation---Lost in translation?}
\bjournal{Int. J. Epidemiol.}
\bvolume{173}
\bpages{739--742}.
\bptok{imsref}%
\end{barticle}
%
\endbibitem

\bibitem[\protect\citeauthoryear{von
{B}ortkiewicz}{1904}]{von-Bortkiewicz-1904}
%
\begin{barticle}[auto:STB|2013/12/09|07:59:19]
\bauthor{\bparticle{von} \bsnm{{B}ortkiewicz},~\bfnm{L.}\binits{L.}}
(\byear{1904}).
\btitle{\"Uber die Methode der ``standard population''}.
\bjournal{Bull. Int. Stat. Inst.}
\bvolume{14}
\bpages{417--437}.
\bptok{imsref}%
\end{barticle}
%
\endbibitem

\bibitem[\protect\citeauthoryear{Walker and Duncan}{1967}]{Walker-Duncan-1967}
%
\begin{barticle}[mr]
\bauthor{\bsnm{Walker},~\bfnm{Strother~H.}\binits{S.~H.}} \AND
\bauthor{\bsnm{Duncan},~\bfnm{David~B.}\binits{D.~B.}}
(\byear{1967}).
\btitle{Estimation of the probability of an event as a function of several
independent variables}.
\bjournal{Biometrika}
\bvolume{54}
\bpages{167--179}.
\bid{issn={0006-3444}, mr={0217928}}
\bptok{imsref}%
\end{barticle}
%
\endbibitem

\bibitem[\protect\citeauthoryear{Westergaard}{1882}]{Westergaard-1882}
%
\begin{bbook}[auto:STB|2013/12/09|07:59:19]
\bauthor{\bsnm{Westergaard},~\bfnm{H.}\binits{H.}}
(\byear{1882}).
\btitle{Die Lehre von der Mortalit\"at und Morbilit\"at}.
\bpublisher{Fischer}, \blocation{Jena}.
\bptok{imsref}%
\end{bbook}
%
\endbibitem

\bibitem[\protect\citeauthoryear{Westergaard}{1901}]{Westergaard-1901}
%
\begin{bbook}[auto:STB|2013/12/09|07:59:19]
\bauthor{\bsnm{Westergaard},~\bfnm{H.}\binits{H.}}
(\byear{1901}).
\btitle{Die Lehre von der Mortalit\"at und Morbilit\"at},
\bedition{2nd} ed.
\bpublisher{Fischer}, \blocation{Jena}.
\bptok{imsref}%
\end{bbook}
%
\endbibitem

\bibitem[\protect\citeauthoryear{Westergaard}{1916}]{Westergaard-1916}
%
\begin{barticle}[auto:STB|2013/12/09|07:59:19]
\bauthor{\bsnm{Westergaard},~\bfnm{H.}\binits{H.}}
(\byear{1916}).
\btitle{Scope and method of statistics}.
\bjournal{Pub. Am. Stat. Ass.}
\bvolume{15}
\bpages{229--276}.
\bptok{imsref}%
\end{barticle}
%
\endbibitem

\bibitem[\protect\citeauthoryear{Westergaard}{1918}]{Westergaard-1918}
%
\begin{barticle}[auto:STB|2013/12/09|07:59:19]
\bauthor{\bsnm{Westergaard},~\bfnm{H.}\binits{H.}}
(\byear{1918}).
\btitle{On the future of statistics}.
\bjournal{J. Roy. Stat. Soc.}
\bvolume{81}
\bpages{499--520}.
\bptok{imsref}%
\end{barticle}
%
\endbibitem

\bibitem[\protect\citeauthoryear{Woodbury}{1922}]{Woodbury-1922}
%
\begin{barticle}[auto:STB|2013/12/09|07:59:19]
\bauthor{\bsnm{Woodbury},~\bfnm{R.~M.}\binits{R.~M.}}
(\byear{1922}).
\btitle{Westergaard's method of expected deaths as applied to the study of
infant mortality}.
\bjournal{J. Amer. Statist. Assoc.}
\bvolume{18}
\bpages{366--376}.
\bptok{imsref}%
\end{barticle}
%
\endbibitem

\bibitem[\protect\citeauthoryear{Xie}{1989}]{Xie-1989}
%
\begin{barticle}[auto:STB|2013/12/09|07:59:19]
\bauthor{\bsnm{Xie},~\bfnm{Y.}\binits{Y.}}
(\byear{1989}).
\btitle{An alternative purging method---Controlling the composition-dependent
interaction in an analysis of rates}.
\bjournal{Demography}
\bvolume{26}
\bpages{711--716}.
\bptok{imsref}%
\end{barticle}
%
\endbibitem

\bibitem[\protect\citeauthoryear{Yamaguchi}{2011}]{Yamaguchi-2011}
%
\begin{barticle}[auto:STB|2013/12/09|07:59:19]
\bauthor{\bsnm{Yamaguchi},~\bfnm{K.}\binits{K.}}
(\byear{2011}).
\btitle{Decomposition of inequality among groups by counterfactual
modeling: An
analysis of the gender wage gap in Japan}.
\bjournal{Sociol. Methodol.}
\bvolume{41}
\bpages{223--255}.
\bptok{imsref}%
\end{barticle}
%
\endbibitem

\bibitem[\protect\citeauthoryear{Yule}{1900}]{Yule-1900}
%
\begin{barticle}[auto:STB|2013/12/09|07:59:19]
\bauthor{\bsnm{Yule},~\bfnm{G.~U.}\binits{G.~U.}}
(\byear{1900}).
\btitle{On the association of attributes in statistics}.
\bjournal{Philos. Trans. R. Soc. Lond. Ser. A Math. Phys. Eng. Sci.}
\bvolume{194}
\bpages{257--319}.
\bptok{imsref}%
\end{barticle}
%
\endbibitem

\bibitem[\protect\citeauthoryear{Yule}{1903}]{Yule-1903}
%
\begin{barticle}[auto:STB|2013/12/09|07:59:19]
\bauthor{\bsnm{Yule},~\bfnm{G.~U.}\binits{G.~U.}}
(\byear{1903}).
\btitle{Notes on the theory of association of attributes in statistics}.
\bjournal{Biometrika}
\bvolume{2}
\bpages{121--134}.
\bptok{imsref}%
\end{barticle}
%
\endbibitem

\bibitem[\protect\citeauthoryear{Yule}{1906}]{Yule-1906}
%
\begin{barticle}[auto:STB|2013/12/09|07:59:19]
\bauthor{\bsnm{Yule},~\bfnm{G.~U.}\binits{G.~U.}}
(\byear{1906}).
\btitle{On the changes in the marriage- and birth-rates in England and Wales
during the past half century; with an inquiry as to their probable causes}.
\bjournal{J. Roy. Stat. Soc.}
\bvolume{69}
\bpages{88--147}.
\bptok{imsref}%
\end{barticle}
%
\endbibitem

\bibitem[\protect\citeauthoryear{Yule}{1911}]{Yule-1911}
%
\begin{bbook}[auto:STB|2013/12/09|07:59:19]
\bauthor{\bsnm{Yule},~\bfnm{G.~U.}\binits{G.~U.}}
(\byear{1911}).
\btitle{An Introduction to the Theory of Statistics}.
\bpublisher{Griffin}, \blocation{London}.
\bptok{imsref}%
\end{bbook}
%
\endbibitem

\bibitem[\protect\citeauthoryear{Yule}{1912}]{Yule-1912}
%
\begin{barticle}[auto:STB|2013/12/09|07:59:19]
\bauthor{\bsnm{Yule},~\bfnm{G.~U.}\binits{G.~U.}}
(\byear{1912}).
\btitle{On the methods of measuring association between two attributes}.
\bjournal{J. Roy. Stat. Soc.}
\bvolume{75}
\bpages{579--652}.
\bptok{imsref}%
\end{barticle}
%
\endbibitem

\bibitem[\protect\citeauthoryear{Yule}{1920}]{Yule-1920}
%
\begin{bbook}[auto:STB|2013/12/09|07:59:19]
\bauthor{\bsnm{Yule},~\bfnm{G.~U.}\binits{G.~U.}}
(\byear{1920}).
\btitle{The Fall of the Birth-Rate}.
\bpublisher{Cambridge Univ. Press}, \blocation{Cambridge}.
\bptok{imsref}%
\end{bbook}
%
\endbibitem

\bibitem[\protect\citeauthoryear{Yule}{1934}]{Yule-1934}
%
\begin{barticle}[auto:STB|2013/12/09|07:59:19]
\bauthor{\bsnm{Yule},~\bfnm{G.~U.}\binits{G.~U.}}
(\byear{1934}).
\btitle{On some points relating to vital statistics, more especially statistics
of occupational mortality (with discussion)}.
\bjournal{J. Roy. Stat. Soc.}
\bvolume{97}
\bpages{1--84}.
\bptok{imsref}%
\end{barticle}
%
\endbibitem

\end{thebibliography}
\end{document}